\documentclass[twocolumn,aps,showpacs,superscriptaddress,longbibliography]{revtex4-1}

\usepackage{graphicx}
\usepackage[normalem]{ulem}
\usepackage{color}
\usepackage{natbib}
\usepackage{latexsym}

\usepackage[utf8]{inputenc}

\begin{document}

\title{Ideal pairing of the Stokes and anti-Stokes photons in the Raman process}

\author{Kishore Thapliyal}
\email{kishore.thapliyal@upol.cz} \affiliation{Joint Laboratory of Optics of
Palack\'{y} University and Institute of Physics of CAS, Faculty of Science,
Palack\'{y} University, 17. listopadu 12, 771 46 Olomouc, Czech Republic}

\author{Jan Pe\v{r}ina,~Jr.}
\email{jan.perina.jr@upol.cz} \affiliation{Joint Laboratory of Optics of
Palack\'{y} University and Institute of Physics of CAS, Institute of Physics,
Czech Academy of Sciences, 17. listopadu 50a, 771 46 Olomouc, Czech Republic}

\begin{abstract}
A quantum model of the Raman process with the independent Stokes and
anti-Stokes nonlinear interactions is developed to study nonclassical
correlations between the photons in the Stokes and anti-Stokes fields. The role
of the laser pump amplitude, the ratio of the Stokes and anti-Stokes coupling
constants and the population and losses of the vibrational mode in forming the
correlations is elucidated. The $ g^{(2)} $ intensity cross-correlation
function, noise-reduction-factor, two-mode principal squeezing variance,
logarithmic negativity, non-classicality depth, steering parameter and the Bell
parameter are analyzed side-by-side to shed light to the correlations between
the Stokes and anti-Stokes fields. Conditions for having the Stokes and
anti-Stokes fields composed of only photon pairs, similarly as it occurs in
twin beams in parametric down-conversion, are revealed. They allow for nonzero
mean thermal phonon numbers.
\end{abstract}

\maketitle

\section{Introduction}

With the recent development in various areas of quantum technology
\cite{dowling2003quantum} search for new physical systems or processes able to
generate useful quantum states has escalated. These quantum states are endowed
with specific properties not available when only classical states are
considered. Negative values of the Glauber-Sudarshan phase-space
quasi-distribution $P$ \cite{glauber1963coherent,sudarshan1963equivalence}
represent their most important feature from which their specific properties
originate. Entangled states with genuine quantum correlations represent the
most important class of such states. They cannot be written as a product of
states of each subsystem. This means that the entangled states exhibit nonlocal
correlations. In their simplest form of entangled photon pairs, they have been
found useful, e.g., in quantum teleportation \cite{bennett1993teleporting},
dense coding \cite{bennett1992communication}, metrology
\cite{giovannetti2011advances} and cryptography \cite{ekert1991quantum}. Even
stronger quantum correlations between two photons, known as the
Einstein-Podolsky-Rosen (EPR) steering and the Bell non-locality, exist and
they have been applied, e.g., in device-independent quantum cryptography
\cite{branciard2012one,acin2006bell}. Also the quantum states with greater
numbers of photons, exhibiting correlations in photon numbers have been studied
recently using photon-number-resolving detection (\cite{perina2016coherent} and
references therein).

The process of parametric down-conversion is traditionally used to generate
various kinds of quantum (entangled) states \cite{Mandel1995}. Nevertheless,
other nonlinear processes, like the Raman process \cite{raman1928new} or the
four-wave mixing \cite{Boyd2003}, are also endowed with the capability to
generate quantum states \cite{perina1992quantum}. On one side these third-order
nonlinear processes are weaker than parametric down-conversion. On the other
side, they are more complex as they involve four interacting fields and so they
are more flexible in the generation of quantum states. Here, we concentrate our
attention to the Raman process, in which the Stokes and anti-Stokes photons are
generated, similarly as the signal and idler photons emerge in parametric
down-conversion. However, there is a principal difference in the microscopic
nature of both processes. Whereas the signal and idler photons emerge as a
photon pair in one quantum event, the Stokes and anti-Stokes photons originate
in two quantum events, the first in the Stokes interaction and the second in
the anti-Stokes interaction of the whole Raman process
\cite{perina1992quantum,PerinaJr1997}. The quantum correlations between the
Stokes and anti-Stokes photons are mediated by the phonons of the vibrational
mode that participates in both interactions. The correlations are established
by phonons that are produced in the Stokes interaction together with the Stokes
photons and later disappear in the anti-Stokes interaction leaving the
anti-Stokes photons. Owing to this microscopic mechanism, the correlations
between the Stokes and anti-Stokes modes are influenced by detailed conditions
of the Raman process and, as a consequence, these correlations do not have to
be strong, or even quantum. On the other hand, the more complex microscopic
mechanism of the Raman process allows us to obtain states not attainable in
parametric down-conversion.

Correlations between the Stokes and anti-Stokes modes in the Raman process have
been theoretically studied in \cite{perina1992quantum,PerinaJr1997}. The
quantum theory of the Raman process was introduced in \cite{walls1970quantum}
and the generation of several types of the nonclassical states was discussed
over the time (for the review, see
\cite{perina1991quantum,miranowicz1994quantum}). Recently, the correlations
between the optical and phonon modes both at resonance and out-of resonance, as
well as the generalization of the Raman process to the hyper-Raman process have
been investigated
\cite{thapliyal2019nonclassicality,thapliyal2019lower,thapliyal2020quasidistribution}.
The fields in the Raman process were experimentally investigated in inelastic
scattering of the laser light from water \cite{kasperczyk2016temporal}, Rb
vapors \cite{podhora2017nonclassical} and diamond \cite{anderson2018two}. The
light scattered from diamond exhibited the Bell nonlocal correlations
\cite{velez2019time}.

The Raman process is also investigated under the conditions in which we have a
large number of non-resonant weakly-interacting  vibrational modes, instead of
one strongly-interacting resonant vibrational mode. In this case the
correlations in the Stokes and anti-Stokes fields emerge due to the exchange of
virtual phonons \cite{de2020lifetime}. Successful theory describing this
process is based upon the unitary transformation that introduces photon pairs
as quasi-particles, in close analogy with the transformation that reveals the
Cooper pairs in superconductivity \cite{saraiva2017photonic,de2019stokes}. As
there occurs no distinct vibrational frequency in the interaction, the spectra
of the emitted Stokes and anti-Stokes fields are broad without peaks. The
Stokes and anti-Stokes photons are ideally paired and there occurs the
entanglement in their frequencies. Such Raman process exhibits many features
typical of four-wave mixing \cite{Boyd2003}. It has been extensively studied in
Refs.~\cite{saraiva2017photonic,de2019stokes,de2020lifetime}.

We note that the Raman process in the form of inelastic scattering (the Raman
scattering) of photons from molecules, has been used to build a quantum memory
\cite{dou2018broadband,ding2018raman,jing2019entanglement}, perform long
distance quantum communication \cite{duan2001long}, quantum state transfer
between the light and matter \cite{matsukevich2004quantum}, entangle remote
super-conducting circuits \cite{campagne2018deterministic}, generate single
photons \cite{chen2006deterministic}, photon-phonon correlated pairs
\cite{riedinger2016non} and other nonclassical states
\cite{meekhof1996generation,kuzmich2003generation,lee2012macroscopic,velez2019preparation}.
Also, the Raman scattering is frequently observed as an unwanted source of the
noise, e.g., in the wavelength-multiplexed quantum cryptographic systems
\cite{eriksson2019wavelength} or in photon-pair generation in optical fibers
via the four-wave mixing \cite{Li2005,Fulconis2005,Fan2005}.

Being inspired by the above applications of the Raman process in quantum
information tasks, we study quantum as well as classical correlations between
the Stokes and anti-Stokes modes using the quantum model of the Raman process
with one distinct vibrational mode and the strong coherent pump beam. Using the
appropriate nonlinear momentum operator that involves the independent Stokes
and anti-Stokes interactions we derive the corresponding Heisenberg equations.
Their solution then allows us to determine the field characteristics including
the intensity cross-correlation functions, the logarithmic negativity, the EPR
steering parameters and the Bell parameter under typical conditions. They are
compared with those characterizing ideal twin beams composed only of photon
pairs and originating in parametric down-conversion under the conditions
similar to those in the Raman process. In the Raman process, among others, we
address the role of damping of the vibrational mode and the amount of thermal
phonons present naturally in this mode.

The paper is organized as follows. We introduce the quantum model of the Raman
process and solve its dynamics in Sec.~\ref{sec:Model-and-solution}. Various
parameters quantifying two-mode correlations are discussed in
Sec.~\ref{sec:math}. A simplified model of the Raman process is analytically
treated in Sec.~\ref{sec:max} considering the initial vacuum vibrational mode.
The model elucidating the role of non-zero initial mean phonon numbers is
presented in Sec.~\ref{sec:nzero-nv}. The general discussion that takes into
account the phonon losses and thermal reservoir phonons is given in
Sec.~\ref{sec:nzero-nt}. Conclusions are drawn in Sec.~\ref{sec:conc}.
\begin{figure}  
  \includegraphics[width=0.99\hsize]{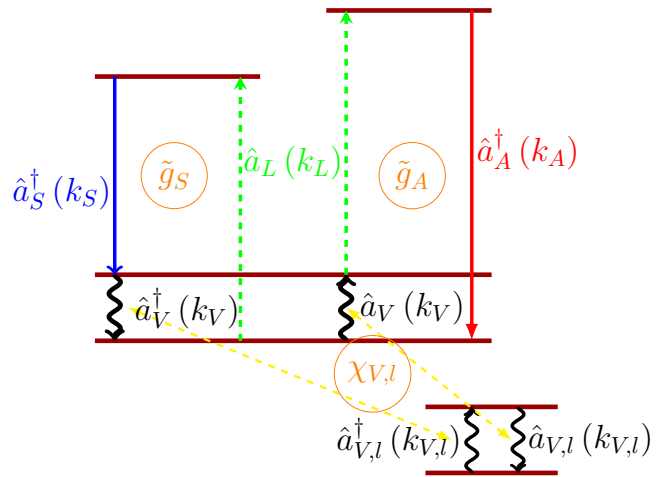}
  \caption{Schematic diagram of the model of the Raman process: A photon in pump mode with
  wave vector $ k_L $ with annihilation operator $ \hat{a}_{L} $ is either converted
  into a phonon with wave vector $ k_V $ and annihilation operator $ \hat{a}_{V} $ and
  a Stokes photon with wave vector $ k_S $ and annihilation operator $ \hat{a}_{S}$ (Stokes interaction with coupling constant
  $ \tilde{g}_{S} $) or is annihilated together with a phonon to give raise to
  an anti-Stokes photon with wave vector $ k_A $ and annihilation operator $ \hat{a}_{A} $
  (anti-Stokes interaction with coupling constant $ \tilde{g}_{A} $). Inversion
  processes also occur. The vibrational mode with phonons are further damped via
  their interaction with reservoir phonon modes with wave vectors $ k_{V,l} $
  and annihilation operators $ \hat{a}_{V,l} $ (linear coupling constants $
  \chi_{V,l} $).}
\label{fig1}
\end{figure}

\section{Quantum model of the Raman process\label{sec:Model-and-solution}}

We consider the monochromatic waves in the laser (frequency $ \omega_L $, wave
vector $ k_L $), Stokes ($ \omega_S $, $ k_S $), anti-Stokes ($ \omega_A $, $
k_A $), and vibration ($ \omega_V $, $ k_V $) modes that propagate along a
nonlinear crystal of length $ L $ and mutually interact in the Raman process
described in the scheme of Fig.~\ref{fig1}. Moreover, the laser pump beam is
assumed to have a strong classical amplitude $ |\alpha_L|\exp(i\phi_L) $ which
results in the following nonlinear momentum operator $ \hat{G}(z) $:
\begin{eqnarray}  
 \hat{G}(z) &=& \hbar k_{S}\hat{a}_{S}^{\dagger}(z)\hat{a}_{S}(z)
  + \hbar k_{A}\hat{a}_{A}^{\dagger}(z)\hat{a}_{A}(z)  \nonumber \\
 & & + \hbar k_{V}\hat{a}_{V}^{\dagger}(z)\hat{a}_{V}(z)  \nonumber \\
 & & \hspace{-5mm} + \Bigl[\hbar \tilde{g}_{S} \hat{a}_{V}^{\dagger}(z)\hat{a}_{S}^{\dagger}(z)|\alpha_L|\exp\left(ik_L z+i\phi_L\right)
   \nonumber \\
 & & \hspace{-5mm} + \hbar \tilde{g}_A\hat{a}_{V}(z)\hat{a}_{A}^{\dagger}(z)|\alpha_L|\exp\left(ik_{L}z+i\phi_{L}\right)+\textrm{H.c.}\Bigr].
\label{eq:Ham}
\end{eqnarray}
In Eq.~(\ref{eq:Ham}), the symbols $ \hat{a}_{S} $, $ \hat{a}_{A} $ and $
\hat{a}_{V} $ ($ \hat{a}_{S}^\dagger $, $ \hat{a}_{A}^\dagger $ and $
\hat{a}_{V}^\dagger $) stand, in turn, for the annihilation (creation)
operators of the Stokes, anti-Stokes and vibrational modes. The nonlinear
coupling constant in the Stokes (anti-Stokes) interaction is denoted as $
\tilde{g}_S $ ($ \tilde{g}_A $), $ \hbar $ is the reduced Planck constant and
H.c. replaces the Hermitian conjugated terms. The law of energy conservation
requires the following conditions for the mode frequencies:
\begin{equation}  
 \omega_{S}=\omega_{L}-\omega_{V}, \hspace{2mm} \omega_{A}=\omega_{L}+\omega_{V}.
\label{2}
\end{equation}

The correlations between the Stokes and anti-Stokes modes emerge via the
interaction with the common vibrational mode whose dynamics plays an important
role. For this reason, we describe both its damping and population by thermal
phonons. Damping of the vibrational mode originates in its interaction with the
reservoir composed of a large number of vibrational modes with frequency $
\omega_V $, wave vectors $ k_{V,l} $ and linear coupling constants $ \chi_{V,l}
$. Introducing their annihilation (creation) operators $ \hat{a}_{V,l} $ ($
\hat{a}_{V,l}^\dagger $), we can write the appropriate interaction momentum
operator $ \hat{G}_R(z) $ as follows:
\begin{eqnarray}  
 \hat{G}_{R}(z) &=& \hbar\sum_{l} k_{V,l}\hat{a}_{V,l}^{\dagger}(z)\hat{a}_{V,l}(z)
  \nonumber \\
  & &  - \hbar\sum_l \left[
  \chi_{V,l}\hat{a}_{V}^{\dagger}(z)\hat{a}_{V,l}(z)+\textrm{H.c.}\right].
\label{eq:Ham-res}
\end{eqnarray}

Spatial evolution of the optical fields and vibrational modes governed by the
overall momentum operator $ \hat{G}(z) + \hat{G}_{R}(z) $ is described by the
Heisenberg equations analogous to those commonly written for the Hamiltonian, $
d\hat{a}(z)/dz = -i/\hbar [\hat{G}(z),\hat{a}(z)] $. { This corresponds to
the scheme of field photon-flux quantization \cite{Huttner1990} suitable also
for nonhomogeneous media.} We note that this method has been widely applied to
study the propagation of nonlinearly interacting optical fields (see
\cite{perina2016coherent,perina1991quantum,PerinaJr2000} and references
therein). In our case, the Heisenberg equations corresponding to the momentum
operator $ \hat{G}(z) + \hat{G}_R(z) $ take the form:
\begin{eqnarray} 
 \frac{d\hat{a}_{S}(z)}{dz} & = & ik_{S}\hat{a}_{S}(z) + g_{S}\hat{a}_{V}^{\dagger}(z)\exp\left(ik_{L}z\right),\nonumber \\
 \frac{d\hat{a}_{A}(z)}{dz} & = & ik_{A}\hat{a}_{A}(z) + g_{A}\hat{a}_{V}(z)\exp\left(ik_{L}z\right),\nonumber \\
 \frac{d\hat{a}_{V}(z)}{dz} & = & ik_{V}\hat{a}_{V}(z) + g_{S}\hat{a}_{S}^{\dagger}(z)\exp\left(ik_{L}z\right) \nonumber\\
 & & -g_{A}^{*}\hat{a}_{A}(z)\exp\left(-ik_{L}z\right) - i \sum_l \chi_{V,l}\hat{a}_{V,l}(z),\nonumber\\
 \frac{d\hat{a}_{V,l}(z)}{dz} & =& ik_{V,l}\hat{a}_{V,l}(z) - i\chi_{V,l}^{*}\hat{a}_{V}(z)
\label{eq:eq-mot}
\end{eqnarray}
and $ g_{S,A} =  i\tilde{g}_{S,A}|\alpha_L|\exp(i\phi_{L}) $.

The equations in (\ref{eq:eq-mot}) were derived assuming the following
canonical commutation relations among the field operators at arbitrary
positions $ z $ and $ z' $ { compatible with photon-flux quantization}:
\begin{eqnarray}   
 { [ \hat{a}_b(z),\hat{a}_c(z')] } &=& 0 , \nonumber \\
 { [ \hat{a}_b(z),\hat{a}_c^\dagger(z')] } &=& \delta_{bc} \delta(z-z') ,
  \hspace{5mm} b,c = S,A,V ;
\label{4a}
\end{eqnarray}
$ \delta_{bc} $ stands for the Kronecker symbol and $ \delta $ means the Dirac
function. The commutation relations (\ref{4a}) represent the spatial analog of
the usual commutation relations written in time { and originating in energy
quantization. The quantization of photon flux is compatible with the problems
of propagating optical fields as it guarantees the continuity of electric-field
amplitude per one quantized photon at the boundaries and as such it can be
applied also to field propagation in nonhomogeneous media. The effects of
spectral and spatial dispersion important for spatially-localized optical
pulses can then be addressed using the spectral and spatial mode decomposition
(see, e.g., in \cite{PerinaJr2019a}). Though the energy and photon-flux
quantization schemes are different, they lead to similar quantum effects. This
can be understood from the very similar forms of the Heisenberg equations
derived for 'unperturbed' monochromatic plane waves in both quantization
schemes (formal substitution $ t \rightarrow z/v $ where $ v $ is an average
velocity of the propagating fields). }

The solution of Eqs.~(\ref{eq:eq-mot}) conserves the following quantity
composed of photon/phonon numbers:
\begin{eqnarray} 
 N &=& \hat{a}_{S}^\dagger(z)\hat{a}_{S}(z) -
  \hat{a}_{A}^\dagger(z)\hat{a}_{A}(z) - \hat{a}_{V}^\dagger(z)\hat{a}_{V}(z)
   \nonumber \\
  & &  - \sum_l \hat{a}_{V,l}^\dagger(z)\hat{a}_{V,l}(z).
\label{5}
\end{eqnarray}
The set of Eqs.~(\ref{eq:eq-mot}) of linear operator differential equations can
be solved following the approach applied in
Ref.~\cite{pieczonkova1981statistical,chizhov2009stokes}. Introducing the interaction picture, in
which the operator envelopes $ \hat{A}_{S,A,V} =
\hat{a}_{S,A,V}\exp(-ik_{S,A,V}z) $ are defined, and assuming the phase matching of the interacting fields ($ k_L - k_V = k_S $, 
$ k_L + k_V = k_A $) we write the solution to
Eqs.~(\ref{eq:eq-mot}) in the form
\begin{eqnarray}  
 \hat{A}_{V}(z) & = & f_{1}(z)\hat{A}_{V}(0)+f_{2,S}(z) \hat{A}_{S}^{\dagger}(0)
   - f_{2,A}^*(z)\hat{A}_{A}(0)  \nonumber \\
   & & + \sum_l f_{1,l}(z)\hat{A}_{V,l}(0),  \nonumber \\
 \hat{A}_{S}(z) & = & f_{2,S}(z)\hat{A}_{V}^{\dagger}(0) + f_{3,S}(z)\hat{A}_{S}(0) +
   f_{4,S}(z)\hat{A}_{A}^{\dagger}(0)  \nonumber \\
   & & + \sum_l f_{2,l}(z)\hat{A}_{V,l}^{\dagger}(0), \nonumber\\
 \hat{A}_{A}(z) & = & f_{2,A}(z) \hat{A}_{V}(0) - f_{4,A}(z)\hat{A}_{S}^{\dagger}(0)
   -f_{3,A}(z) \hat{A}_{A}(0)  \nonumber \\
  & & + \sum_l f_{3,l}(z) \hat{A}_{V,l}(0).
\label{eq:exact-sol}
\end{eqnarray}
The functions introduced in the solution~(\ref{eq:exact-sol}) are defined as
\begin{eqnarray}  
  f_{1}(z) &=& \frac{[\Gamma q(z)-\gamma p(z)]  h(z)}{\Gamma} , \nonumber\\
  f_{2,b}(z) &=& \frac{4ig_{b}p(z) h(z) }{\Gamma},  \nonumber\\
  f_{3,b}(z) &=& \frac{\Gamma|g_{b}|^{2}-|g_{\bar{b}}|^{2}[\Gamma q(z)+\gamma p(z)] h(z) }{\Gamma\Omega^{2}}, \nonumber\\
  f_{4,b}(z) &=& \frac{g_{b}g_{\bar{b}} \{\Gamma- [\Gamma q(z)+\gamma p(z)]h(z)\}
  }{\Gamma\Omega^{2}}.\nonumber \\
  & & (b,\bar{b}) = (S,A)\, \mathrm{or} \,(b,\bar{b}) = (A,S), \nonumber \\
  f_{1,l}(z) &=& -i(\chi_{V,l}/\Gamma) \bigl\{ \Delta k_{V,l}\bigl[ \Gamma g_{l}(z)
   - [\Gamma q(z)-\gamma p(z)] h(z)\bigr] \nonumber \\
   & & + 4i\Omega^{2} p(z) h(z) \bigr\} \left[\Delta k_{V,l}(\gamma-i\Delta k_{V,l}) +i\Omega^{2}\right]^{-1},
   \nonumber \\
  f_{2,l}(z) &=& ig_{S}(\chi_{V,l}^{*}/\Gamma)\Bigl\{\Gamma g_{l}(z) -\bigl[4i\Delta k_{V,l} p(z) + [\Gamma q(z) \nonumber \\
   & &  +\gamma p(z)]\bigr] h(z) \bigr\} \left[ \Delta k_{V,l} (\gamma/2+i\Delta k_{V,l})-i\Omega^{2}\right]^{-1}, \nonumber\\
  f_{3,l}(z) & = & ig_{A}(\chi_{V,l}/\Gamma) \bigl\{ \Gamma g_l(z) + \bigl[4i\Delta k_{V,l} p(z)-[\Gamma q(z) \nonumber \\
   & &  + \gamma p(z)] \bigr] h(z) \bigr\} \left[ \Delta k_{V,l} (\gamma/2-i\Delta
   k_{V,l})+i\Omega^{2}\right]^{-1}. \nonumber \\
  & &
\label{eq:terms}
\end{eqnarray}
In Eqs.~(\ref{eq:terms}), we use functions  $ g_l(z) = \exp(i\Delta k_{V,l}z)
$, $ h(z) = \exp(-\gamma z/4) $, $ p(z) = \sinh(\Gamma z/4) $, $ q(z) =
\cosh(\Gamma z/4) $ and constants $\Omega= \sqrt{|g_{A}|^{2}-|g_{S}|^{2}} $, $
\Gamma = \sqrt{\gamma^{2}-16\Omega^{2}} $ and $ \Delta k_{V,l} = k_{V,l} -
k_{V} $. The damping constant $ \gamma $ is given, according to the
Wigner-Weisskopf theory \cite{perina1992quantum}, as $ \gamma = 2\pi |\chi_V|^2
\varrho_V $, where $ \chi_V $ is an average reservoir coupling constant and $
\varrho_V $ stands for the density of vibrational reservoir modes. We note that
the functions defined in Eqs.~(\ref{eq:terms}) obey several relations that
originate in the bosonic commutation relations at an arbitrary position $ z $.
{ This is so due to the rigorous description of the vibrational reservoir
that causes damping of the vibrational mode and that also acts on the
vibrational mode via the corresponding fluctuating operator forces that
compensate for damping in the quantum evolution (fluctuation--dissipation
theorem).}

In our analysis, we assume that the Stokes and anti-Stokes modes are initially
in the vacuum states, whereas the vibrational mode and its reservoir
vibrational modes are initially in thermal states with the average phonon
numbers $ \langle \hat{n}_{V}(0)\rangle = n_{V} $ and $ \langle \hat{n}_{V,l} \rangle = n_T $, respectively.
The normal characteristic function $ C_{\cal N}(\beta_S,\beta_A;z) $, which is
a commonly used tool to access all useful physical quantities (see
Sec.~\ref{sec:math} below), encompasses the combined Stokes and anti-Stokes
modes and it is defined as
\begin{eqnarray}  
 C_{\cal N}(\beta_S,\beta_A;z) &=& \langle \exp[\beta_S\hat{A}_S^\dagger(z)+
  \beta_A\hat{A}_A^\dagger(z)]  \nonumber \\
 & & \times \exp[-\beta_S^*\hat{A}_S(z)- \beta_A^*\hat{A}_A(z)]\rangle.
\label{8}
\end{eqnarray}
It takes the following Gaussian form for the solution written in
Eqs.~(\ref{eq:exact-sol}) \cite{perina1992quantum,PerinaJr2000}:
\begin{eqnarray}  
 C_{\cal N}(\beta_S,\beta_A;z) &=& \exp[ -B_S(z)|\beta_S|^2 -
  B_A(z)|\beta_A|^2] \nonumber \\
 & & \times \exp[ D_{SA}(z)\beta_S^*\beta_A^* + {\rm c.c.} ]
\label{9}
\end{eqnarray}
and symbol c.c. replaces the complex conjugated term.

In Eq.~(\ref{9}), the $ z $-dependent coefficients are obtained as follows:
\begin{eqnarray}   
 B_{S}(z) &=& b_S(z,z) , \nonumber \\
 b_S(z,z') &=& \langle \hat{A}_{S}^{\dagger}(z)\hat{A}_{S}(z')\rangle
  = f_{2,S}^*(z) f_{2,S}(z') ( n_{V} +1) \nonumber \\
  & & \hspace{-5mm} + f_{4,S}^*(z)f_{4,S}(z') + \sum_l f_{2,l}^*(z) f_{2,l}(z') ( n_T +1), \nonumber \\
 B_{A}(z) &=& b_A(z,z) , \nonumber \\
 b_A(z,z') &=& \langle \hat{A}_{A}^{\dagger}(z)\hat{A}_{A}(z')\rangle =
  f_{2,A}^*(z) f_{2,A}(z')  n_{V}  \nonumber \\
  & & \hspace{-5mm} + f_{4,A}^*(z)f_{4,A}(z') + \sum_l f_{3,l}^*(z) f_{3,l}(z')  n_T ,\nonumber \\
 D_{SA}(z) &=& d_{SA}(z,z) , \nonumber \\
 d_{SA}(z,z') &=& \langle \hat{A}_{S}(z)\hat{A}_{A}(z')\rangle
  = f_{2,S}(z)f_{2,A}(z')  n_{V}  \nonumber \\
  & & \hspace{-5mm} - f_{3,S}(z)f_{4,A}(z') + \sum_l f_{2,l}(z)f_{3,l}(z')
  n_T.
\label{10}
\end{eqnarray}

The terms on the right-hand sides of Eqs.~(\ref{10}) involving the sum over the
vibrational reservoir modes can conveniently be determined from the commutation
relations $ [\hat{A}_S(z),\hat{A}_S^\dagger(z)] = 1 $, $
[\hat{A}_A(z),\hat{A}_A^\dagger(z)] = 1 $, and $ [\hat{A}_S(z),\hat{A}_A(z)] =
0 $ provided that $ z = z' $. In turn, they give us:
\begin{eqnarray}  
 & \sum_l |f_{2,l}(z)|^{2} = -1 - |f_{2,S}(z)|^{2} + |f_{3,S}(z)|^{2} - |f_{4,S}(z)|^{2}, & \nonumber\\
 & \sum_l |f_{3,l}(z)|^{2} = 1 -|f_{2,A}(z)|^{2} - |f_{3,A}(z)|^{2} + |f_{4,A}(z)|^{2}, &  \nonumber \\
 & \sum_l f_{2,l}(z)f_{3,l}(z) = - f_{2,S}(z)f_{2,A}(z) - f_{3,S}(z)f_{4,A}(z) & \nonumber \\
 & \hspace{15mm} + f_{4,S}(z)f_{3,A}(z). &
\label{11}
\end{eqnarray}

\section{Quantification of correlations between the Stokes and anti-Stokes fields \label{sec:math}}

Owing to the relationship between the Stokes and anti-Stokes modes mediated by
the vibrational mode the correlations between the Stokes and anti-Stokes modes
are developing as the Raman process proceeds. These correlations manifest
themselves in various ways starting from the usual classical intensity
correlations and ending with the steering and the Bell nonlocality between the
modes that expresses the exclusively quantum influence of one mode to the other
mode.

The coherence theory defines the normalized intensity cross-correlation
function $ g^{(2)}_{SA}(z,z') $ between the Stokes intensity at position $ z $
and anti-Stokes intensity at position $ z'$
\cite{perina1992quantum,sekatski2012detector},
\begin{equation} 
 g^{(2)}_{SA}(z,z')= \frac{\langle \hat{A}_{S}^{\dagger}(z) \hat{A}_{A}^{\dagger}(z')
  \hat{A}_{A}(z') \hat{A}_{S}(z) \rangle }{ \langle \hat{A}_{S}^{\dagger}(z)\hat{A}_{S}(z)\rangle
  \langle \hat{A}_{A}^{\dagger}(z')\hat{A}_{A}(z')\rangle},
\label{eq:g2}
\end{equation}
to quantify the mutual relation between the Stokes and anti-Stokes fields.
Whereas the mean photon numbers $ \langle \hat{A}_{S}^{\dagger}
\hat{A}_{S}\rangle $ and $ \langle \hat{A}_{A}^{\dagger}\hat{A}_{A}\rangle $ in
Eq.~(\ref{eq:g2}) are given by the coefficients $ B_S $ and $ B_A $ in
Eqs.~(\ref{10}), respectively, the remaining fourth order amplitude correlation
function is determined as follows:
\begin{eqnarray}  
 \langle\hat{A}_{S}^{\dagger}(z)\hat{A}_{A}^{\dagger}(z')\hat{A}_{A}(z')\hat{A}_{S}(z)\rangle
   & & \nonumber \\
  & & \hspace{-20mm} = B_S(z) B_A(z') + |d_{SA}(z,z')|^2.
\label{eq:corr}
\end{eqnarray}
The $ g^{(2)}$ function defined in Eq.~(\ref{eq:g2}) identifies possible
bunching of photons residing in two modes.

Nonclassical character of two-mode fields is often tested by the violation of
the Cauchy-Schwarz inequality that represents one of the simplest and also
most powerful nonclassicality witnesses \cite{sekatski2012detector}:
\begin{eqnarray} 
 &\langle \hat{A}_{S}^{\dagger}(z)\hat{A}_{A}^{\dagger}(z)\hat{A}_{A}(z)
  \hat{A}_{S}(z)\rangle^{2} \le \hspace{1cm} & \nonumber \\
 &\hspace{1cm} \langle \hat{A}_{S}^{\dagger 2}(z) \hat{A}_{S}^{2}(z)\rangle
  \langle \hat{A}_{A}^{\dagger 2}(z) \hat{A}_{A}^{2}(z)\rangle.&
\label{eq:CSI}
\end{eqnarray}
As the Stokes and anti-Stokes modes evolve from the vacuum states, their
statistics remain chaotic and so we have $ \langle \hat{A}_{b}^{\dagger 2}(z)
\hat{A}_{b}^{2}(z)\rangle = 2 [ \langle \hat{A}_{b}^{\dagger}(z)
\hat{A}_{b}(z)\rangle ]^2 $, $ b= S,A $. Under these conditions, the
Cauchy-Schwarz inequality (\ref{eq:CSI}) is transformed into the following
non-classicality inequality:
\begin{equation} 
 g^{(2)}_{SA}(z) > 2.
\label{eq:CSI-res}
\end{equation}

Correlations in photon numbers $ \hat{n}_S \equiv  \hat{a}_S^\dagger
\hat{a}_S =  \hat{A}_S^\dagger \hat{A}_S $ and $ \hat{n}_A
\equiv  \hat{a}_A^\dagger \hat{a}_A  =  \hat{A}_A^\dagger
\hat{A}_A $ of the Stokes and anti-Stokes fields, respectively, are
characterized by the noise-reduction-factor $ R_{SA} $
\cite{perina2016coherent}:
\begin{equation} 
 R_{SA}(z) = \frac{ \langle \left(\Delta [ \hat{n}_S(z)-\hat{n}_A(z) ] \right)^2
  \rangle }{ \langle\hat{n}_S(z)\rangle + \langle\hat{n}_A(z)\rangle };
\label{16}
\end{equation}
$ \Delta \hat{x} = \hat{x} - \langle \hat{x}\rangle $ gives the fluctuation of
operator $ \hat{x} $. The values of $ R_{SA} $ smaller than one are reserved
for nonclassical fields for which the fluctuations of the difference of the
Stokes and anti-Stokes photon numbers are suppressed below the quantum
Poissonian limit. In the limiting case of $ R_{SA}=0 $, the Stokes and
anti-Stokes fields can be considered as composed of only photon pairs having
one photon in the Stokes mode and the accompanying photon in the anti-Stokes
mode. This case represents the most nonclassical two-mode fields. Individual
photons comprising a photon pair are in this case completely indistinguishable
(apart from different frequencies). Provided that the difference in the Stokes
and anti-Stokes field frequencies is omitted, the photons give visibility one
in the Hong-Ou-Mandel interferometer \cite{Friberg1985}, i.e., they coalesce at
a beam splitter. The fraction of paired photons in the two-mode field can
roughly be estimated as $ 1- R_{SA} $. Using the coefficients in
Eqs.~(\ref{10}), the noise-reduction-factor $ R_{SA} $ is expressed as:
\begin{equation} 
 R_{SA}(z) =1+ \frac{ B_S^2(z) + B_A^2(z) - 2| D_{SA}(z)|^2}{ B_S(z) +
  B_A(z)}.
\label{eq:R}
\end{equation}

The Stokes and anti-Stokes fields can also exhibit nonclassical correlations
in their phase properties. The strength of such correlations is described by
the two-mode principal squeezing variance $ \lambda_{SA} $ \cite{Luks1988} that
is determined by the following formula \cite{perina1991quantum}:
\begin{equation} 
 \lambda_{SA}(z)= 1 + B_{S}(z) + B_{A}(z) - 2|D_{SA}(z)|.
\label{eq:R-1}
\end{equation}
We have $ 0<\lambda_{SA}<1 $ for nonclassical fields. The smaller the value of
$ \lambda_{SA} $ is the stronger the quantum phase correlations are and the
more reduced the fluctuations of the relative phase below the quantum limit
given by the Heisenberg uncertainty relations are.

In general, we can quantify the entanglement of the Stokes and anti-Stokes
fields by determining the logarithmic negativity \cite{Hill1997}. It is derived
from the covariance matrix $ \bm{\sigma} $ \cite{adesso2007entanglement},
\begin{eqnarray} 
 & \bm{\sigma}(z) = \left[\begin{array}{cc}
   \bm{\sigma}_{S}(z) & \bm{\sigma}_{SA}(z) \\
   \bm{\sigma}_{SA}(z) & \bm{\sigma}_{A}(z)
   \end{array}\right], &
\label{eq:cov-mat} \\
 & \bm{\sigma}_b(z) = \left[\begin{array}{cc}
   1+2B_b(z) & 0\\
   0 & 1+2B_b(z) \end{array}\right],& b=S,A,\nonumber \\
  & \bm{\sigma}_{SA}(z) = 2\left[\begin{array}{cc}
   {\rm Re}\{D_{SA}(z)\} & -i{\rm Im}\{D_{SA}(z)\} \\
   -i{\rm Im}\{D_{SA}(z)\} & -{\rm Re}\{D_{SA}(z)\} \end{array}\right]. &
\end{eqnarray}
The simplectic eigenvalues $ \xi_{\pm} $ of the partially-transposed covariance
matrix $  \bm{\sigma} $ \cite{Horodecki1996,Peres1996} attain the form
\begin{eqnarray}  
 2\xi_{\pm}^{2}(z) &=& \sum_{b=S,A} [1+2B_{b}(z)]^{2} + 8{\rm Re}\{D_{SA}^{2}(z)\}
   \nonumber \\
 & & \pm 4[1 + B_{S}(z) +B_{A}(z)] \nonumber \\
 & & \times \sqrt{[B_{S}(z)-B_{A}(z)]^{2} + 4{\rm Re}\{D_{SA}^{2}(z)\}}.
\label{20}
\end{eqnarray}
Symbol Re (Im) denotes the real (imaginary) part of an expression. If $
\xi_{-}<1 $ we arrive at the nonzero logarithmic negativity $
E_{\mathcal{N},SA} $ along the formula
\begin{equation} 
 E_{\mathcal{N},SA}(z) = -\ln[\xi_{-}(z)],
\label{22}
\end{equation}
where $ \ln $ stands for the natural logarithm. The entangled states have $
E_{\mathcal{N}}>0 $: The greater the logarithmic negativity $ E_{\mathcal{N}} $
is, the more entangled the state is and the more quantum the state properties
are.

We note that the entropy cannot be applied to quantify the entanglement as the
common state of the Stokes and anti-Stokes fields is in general mixed. Purity $
\mu_{SA}(z) $ that measures the mixedness of a state is determined as
\begin{eqnarray} 
 \mu_{SA}(z) &=& 1/ \sqrt{\det [\bm{\sigma}(z)]} \nonumber \\
  &=& 1 / \bigl[ [1+2B_{S}(z)] [1+2B_{A}(z)] - 4{\rm Re}\{ D_{SA}^2(z) \} \bigr]
   \nonumber \\
  & &
\label{23}
\end{eqnarray}
where $ \det $ denotes the determinant of a matrix. The smaller the purity is
the more complex the internal structure of the state is.

Alternatively, as the reduced states of the Stokes and anti-Stokes modes are
chaotic, i.e. classical, the Lee non-classicality depth $ \tau_{SA} $
\cite{lee1991measure} can be applied to quantify the entanglement between the
Stokes and anti-Stokes fields. It gives the amount of the noise needed to
conceal the non-classicality of the analyzed field. For the considered two-mode
field, the non-classicality depth $ \tau_{SA} $ is determined as follows
\cite{perina2017higher}:
\begin{eqnarray}  
  \tau_{SA}(z) &=& {\rm max} \bigl\{0, - \bigl[B_{S}(z)+B_{A}(z)\bigr]/2  \nonumber \\
  & & \hspace{-3mm} + \sqrt{[B_{S}(z)-B_{A}(z)]^{2}+4|D_{SA}(z)|^{2}}/2 \bigr\}.
\label{eq:non-dep}
\end{eqnarray}
For two-mode Gaussian states, $ \tau_{SA}\le 1/2 $ and the greater the
non-classicality depth $ \tau_{SA} $ is, the more nonclassical the state is.

Owing to the microscopic origin of the correlations, in which the Stokes and
anti-Stokes modes play different roles, the correlations between the modes are
asymmetric. To quantify this asymmetry, we use the steering parameter $
\mathcal{S}_{b\rightarrow c} $ \cite{Cavalcanti2009} that tells us to which
extent we can steer the state of subsystem $ c $ by the measurement on the
state of subsystem $ b $. We have for two-mode Gaussian states and Gaussian
measurements on the state of subsystem $ b $  \cite{kogias2015quantification}:
\begin{eqnarray} 
 \mathcal{S}_{b\rightarrow c}(z) &=& \max\left\{ 0, \ln\bigl(
  \det\bigl[\sigma_b(z)\bigr] /  \det\bigl[\sigma(z)\bigr] \bigr)/2
  \right\}, \nonumber \\
  & &  (b,c) = (S,A)\, {\rm or}\, (A,S).
\label{eq:steer}
\end{eqnarray}

The Stokes and anti-Stokes fields are not only entangled, they may even exhibit
the Bell-type correlations --- the strongest quantum correlations. The
corresponding hidden-variable model
\cite{banaszek1998nonlocality,olivares2004enhancement} is constructed upon the
joint displaced parity operator
\begin{eqnarray} 
 \hat{\Pi}(\beta_S,\beta_A) &=& \hat{D}_{S}(\beta_S)(-1)^{\hat{A}^{\dagger}_S\hat{A}_S}
  \hat{D}_S^{\dagger}(\beta_S) \nonumber \\
 & &  \otimes \hat{D}_A(\beta_A)(-1)^{\hat{A}_A^{\dagger}\hat{A}_A}\hat{D}_A^{\dagger}(\beta_A) ,
\label{eq:parity}
\end{eqnarray}
where the symbol $ \hat{D}_b(\beta_b) $ denotes the displacement operator in mode
$ b $ [$ \hat{D}_b(\beta_b) = \exp(\beta_b \hat{A}_b^\dagger - {\rm H.c.}) $]. The mean value $ \Pi(\beta_S,\beta_A) $ of the operator $
\hat{\Pi}(\beta_S,\beta_A) $ can easily be obtained owing to its relation to
the Wigner function $ W $,
\begin{equation} 
 \Pi(\beta_S,\beta_A) = \frac{\pi^2}{4}W(\beta_S,\beta_A).
\label{eq:exp-par}
\end{equation}
The Wigner function is then determined by the Fourier transform of the
symmetrically-ordered characteristic function $ {\cal C}_{\cal
S}(\beta_S,\beta_A) $ that is derived from its normally-ordered form $ {\cal
C}_{\cal N}(\beta_S,\beta_A) $ in Eq.~(\ref{9}) by the replacement $ B_S
\leftarrow B_S + 1/2 $ and $ B_A \leftarrow B_A + 1/2 $. As the parity operator
is a dichotomic variable, it may be used to define the Bell-like inequalities
\cite{clauser1969proposed}. The corresponding Bell parameter is then defined as
follows using the mean values of suitably chosen displaced parity operators
\cite{banaszek1998nonlocality}:
\begin{eqnarray} 
 \mathcal{B}_{SA}(\beta_{S1},\beta_{S2},\beta_{A1},\beta_{A2}) &=&
  \Pi(\beta_{S1},\beta_{A1})+\Pi(\beta_{S2},\beta_{A1}) \nonumber \\
 & & \hspace{-12mm} \mbox{} +\Pi(\beta_{S1},\beta_{A2}) -\Pi(\beta_{S2},\beta_{A2}).
\label{eq:bell}
\end{eqnarray}
If, for any suitably chosen set of parameters $ \beta_{S1} $, $ \beta_{S2} $, $
\beta_{A1} $ and $ \beta_{A2} $, the Bell parameter $ \mathcal{B}_{SA} $
exceeds two, any hidden-variable theory cannot be considered and we speak about
the nonlocal Bell-type correlations.

\section{The Raman process with an ideal vibrational mode in the initial vacuum state\label{sec:max}}

The strongest correlation between the Stokes and anti-Stokes fields is expected
provided that the vibrational mode is initially in the vacuum state and is not
damped. In this case, there has to exist a Stokes photon for each anti-Stokes
photon because creation of an anti-Stokes photon in the anti-Stokes interaction
is accompanied by annihilation of a vibrational phonon that was created in the
Stokes interaction together with the Stokes photon. Additional thermal phonons,
that exist independently of the Stokes interaction, make the anti-Stokes
interaction independent of the Stokes field and thus weaken the correlations
between the Stokes and anti-Stokes modes. Similarly, damping of the vibrational
mode, that removes the phonons comprising pairs with the Stokes photons and
also adds independent phonons, partially breaks the correlations between the
Stokes and vibrational modes and thus weakens the correlations between the
Stokes and anti-Stokes modes.

In this ideal case, crucial role in forming the correlations is played by the
ratio $ \epsilon $ of the squared moduli of anti-Stokes and Stokes nonlinear coupling constants,
$ \epsilon = |g_{A}|^2/|g_{S}|^2 $. The ratio $ \epsilon $ is a characteristic
parameter of the medium exhibiting the Raman process \cite{parra2016stokes}. In
the analysis of the solution (\ref{eq:exact-sol}), we assume that the phase $
\phi_L $ of the laser pump beam is such that the nonlinear coupling constants $
g_S $ and $ g_A $ are real and positive ($ \phi_L = -\pi/2 $) which maximizes
the quantum correlations in the Raman process. We note that the coupling
constants $ \tilde{g}_S $ and $ \tilde{g}_A $ involved in Eq.~(\ref{eq:Ham})
are real and positive as they quantify the corresponding coupling momentum
(energy). Then we identify two different regimes in the dynamics of the Raman
process as described by the solution (\ref{eq:exact-sol}): The intensities of
the Stokes and anti-Stokes modes exponentially increase for $ \epsilon \le 1 $,
whereas they show periodic solution for $ \epsilon > 1 $.

The validity of the exponential solution is limited by the assumption of
un-depleted pump field that allows the transfer of just a small fraction of the
pump energy into the Stokes and anti-Stokes fields. On the other hand, the
periodic solution originates in the competition of the Stokes and anti-Stokes
interactions for the vibrational phonons and such interactions are known to
behave as if they are phase-mismatched \cite{PerinaJr2000}: The greater the
ratio $ \epsilon > 1 $ is, the greater the effective phase-mismatch is and so
the smaller the fraction of the pump energy accessible for the transfer into
the Stokes and anti-Stokes fields \cite{Boyd2003}. Experimental investigations
of the Raman process under the condition $ \epsilon > 1 $ can be found, e.g.,
in Refs.~\cite{jorio2014optical,kasperczyk2015stokes}. We note that some
aspects of pump depletion in quantum models of the Raman process were
discussed, e.g., in Refs.~\cite{Perina1991,perina1992quantum}.

Straightforward calculations in the case $ \epsilon \le 1 $ leave us with the
following coefficients in the characteristic function $ C_{\cal
N}(\beta_S,\beta_A;z) $ in Eq.~(\ref{9}):
\begin{eqnarray}  
 B_{A}^{\rm id,e}(z) &=& 4\epsilon\sinh^{4}\left(g_{S}z\sqrt{1-\epsilon}/2  \right) /\left(1-\epsilon\right)^{2},
  \nonumber \\
 B_{S}^{\rm id,e}(z) &=& B_{A}^{\rm id,e}(z) + \sinh^{2}\left(g_{S}z\sqrt{1-\epsilon} \right)/ \left( 1-\epsilon \right),
  \nonumber \\
 D_{SA}^{\rm id,e}(z) &=& \sqrt{\epsilon}\left[\epsilon-\cosh\left(g_{S}z\sqrt{1-\epsilon} \right)\right]
  \nonumber \\
 & & \times  \left[\cosh\left(g_{S}z\sqrt{1-\epsilon}\right)-1\right] / \left(1-\epsilon\right)^{2}.
\label{29}
\end{eqnarray}

On the other hand, the coefficients of the characteristic function $ C_{\cal N}
$ for the periodic solution valid for $ \epsilon
> 1 $ are written as:
\begin{eqnarray}  
 B_{A}^{\rm id,p}(z) &=& 4\epsilon\sin^{4}\left(g_{S}z\sqrt{\epsilon-1}/2 \right) /\left(\epsilon-1\right)^{2},
  \nonumber \\
 B_{S}^{\rm id,p}(z) &=& B_{A}^{\rm id,p}(z) + \sin^{2}\left(g_{S}z\sqrt{\epsilon-1} \right)/ \left(\epsilon-1 \right),
  \nonumber \\
 D_{SA}^{\rm id,p}(z) &=& \sqrt{\epsilon}\left[\epsilon-\cos\left(g_{S}z\sqrt{\epsilon-1} \right)\right]
  \nonumber \\
 & & \times  \left[\cos\left(g_{S}z\sqrt{\epsilon-1}\right)-1\right] / \left(\epsilon-1\right)^{2}.
\label{eq:noise-lossless}
\end{eqnarray}

An important characteristic of the coefficients in Eqs.~(\ref{29}) and
(\ref{eq:noise-lossless}), that give the fields properties, is that they depend
on the product of the Stokes nonlinear coupling constant $ \tilde{g}_S $, the
pump-beam amplitude $ |\alpha_L| $ and the interaction length $ z $. For this
reason, we further discuss the fields properties as they depend on the
'cumulative' normalized nonlinear pump amplitude $ |\alpha_L^{\rm n}| $ defined
for a Raman medium of length $ L $:
\begin{equation}  
 |\alpha_L^{\rm n}| \equiv \tilde{g}_S |\alpha_L| L .
\label{31}
\end{equation}
We also introduce the normalized damping constant $ \gamma^{\rm n} \equiv
\gamma L $.

\begin{figure}  
  \includegraphics[width=0.99\hsize]{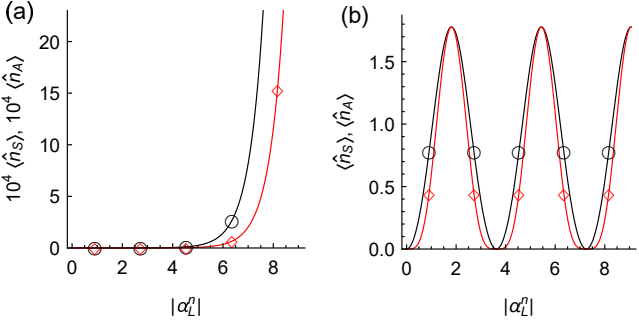}
  \caption{(a,b) Stokes [anti-Stokes] mean photon number $ \langle \hat{n}_S \rangle $
  [$ \langle \hat{n}_A \rangle $] ($ \circ $) [(red $ \diamond $)] in exponential (a) and
  oscillatory (b) regimes.
  The used parameters are: $ \epsilon = 1/4 $ ($ \epsilon = 4 $) in exponential
  (oscillatory) regime, $ \gamma^{\rm n} = 0 $, $ n_V =
  n_T = 0 $.}
\label{fig2}
\end{figure}

The conservation law in Eq.~(\ref{5}) for mean photon/phonon numbers guarantees
that the number $ \langle \hat{n}_A\rangle^{\rm id} $ of anti-Stokes photons
cannot exceed the number $ \langle \hat{n}_S\rangle^{\rm id} $ of Stokes
photons. Whereas the difference $ \langle \hat{n}_S\rangle^{\rm id} - \langle
\hat{n}_A\rangle^{\rm id} $ of these photon numbers, that in fact gives the
number $ \langle \hat{n}_V\rangle^{\rm id} $ of vibrational phonons,
monotonously increases with the increasing pump amplitude $ |\alpha_L^{\rm n}|
$ in the exponential regime [see Fig.~\ref{fig2}(a)], it equals to zero for
specific values of the pump amplitude $ |\alpha_L^{\rm n}| $ in the oscillatory
regime [see Fig.~\ref{fig2}(b)]. For these pump amplitudes numbered by $ m $,
the Stokes and anti-Stokes photon numbers coincide [$ \langle
\hat{n}_S\rangle^{\rm id}(|\alpha_{L,m}^{\rm n}|) = \langle
\hat{n}_A\rangle^{\rm id}(|\alpha_{L,m}^{\rm n}|) $] and moreover the
vibrational mode is in the vacuum state [$ \langle \hat{n}_V\rangle^{\rm
id}(|\alpha_{L,m}^{\rm n}|) =0 $].
\begin{figure}  
  \includegraphics[width=0.87\hsize]{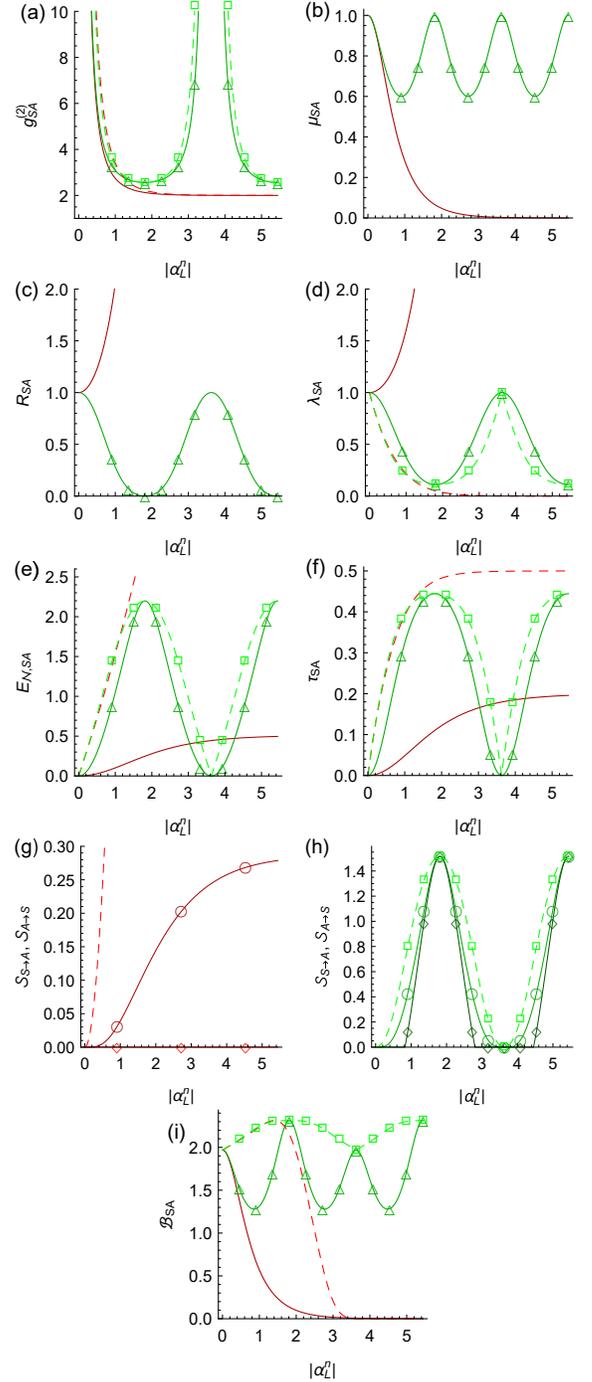}
 \caption{(a) $ g_{SA}^{(2)} $ intensity function, (b) purity
  $ \mu_{SA} $, (c) noise-reduction-factor $ R_{SA} $, (d) principal squeezing
  variance $ \lambda_{SA} $, (e) logarithmic negativity $ E_{{\cal N},SA} $, (f)
  non-classicality depth $ \tau_{SA} $, (g,h) steering parameter
  $ \mathcal{S}_{S\rightarrow A} $ [$ \mathcal{S}_{A\rightarrow S} $] ($ \circ $) [($ \diamond $)]
  in exponential (g) and oscillatory (h) regimes and (i) the Bell parameter $
  {\cal B}_{SA} $ as they depend on normalized
  pump amplitude $ |\alpha_{L}^{\rm n}| $. In (a)---(f) and (i), curves for exponential
  (dark red plain solid curve) and oscillatory (green plain solid curves with $ \triangle $) regimes are plotted.
  In (a)---(i), the light red plain dashed curves and the light green dashed curves with $ \Box $ characterize parametric down-conversion
  with mean signal/idler photon number $ \langle \hat{n}_{\rm spdc}\rangle = ( \langle \hat{n}_S \rangle +
  \langle \hat{n}_A \rangle ) /2 $ in the exponential and oscillatory regimes, respectively
  (for details, see Appendix \ref{sec:app}).
  The used parameters are given in the caption to Fig.~\ref{fig2}.}
\label{fig3}
\end{figure}

In both the exponential and oscillatory regimes, the $ g_{SA}^{(2)} $ intensity
function decreases from high nonclassical values with the increasing pump
amplitude $ |\alpha_L^{\rm n}| $ [see Fig.~\ref{fig3}(a)]. This decrease is
exponential in the exponential regime and ends up at the classical value $
g_{SA}^{(2)} = 2 $. In the oscillatory regime, it stops at the pump amplitude $
|\alpha_{L,1}^{\rm n}| = \pi/\sqrt{\epsilon -1} $ at which the Stokes and
anti-Stokes photon numbers coincide and reach their maximal values. Then,
according to the formulas in Eqs.~(\ref{eq:noise-lossless}) the dependence $
g_{SA}^{(2)}(|\alpha_L^{\rm n}|) $ is periodic and the attained values are
nonclassical for all pump amplitudes. Qualitative difference in the fields
evolution in both regimes is clearly documented by the values of the purity $
\mu_{SA} $. Whereas the purity $ \mu_{SA} $ monotonously decreases from the
initial value $ \mu_{SA} = 1 $ of the pure vacuum state with the increasing
pump amplitude $ |\alpha_L^{\rm n}| $ in the exponential regime [see Fig.~\ref{fig3}(b)], it behaves
periodically in the oscillatory regime. This means that the fields purity $
\mu_{SA} $ initially decreases with the increasing pump amplitude $
|\alpha_L^{\rm n}| $ but then it starts to increase and the fields state is
pure at the pump amplitude $ |\alpha_{L,1}^{\rm n}| $.

In the exponential regime, nonclassical correlations in the Stokes and
anti-Stokes modes as identified by the noise-reduction-factor $ R_{SA} $ are
not built [see Fig.~\ref{fig3}(c)]. On the other hand, they exist in the
oscillatory regime for any pump amplitude $ |\alpha_L^{\rm n}| $ and even reach
their maximal nonclassical value {($  R_{SA}=0 $)} for the pump amplitude $
|\alpha_{L,1}^{\rm n}| $. The correlations in the phases of the Stokes and
anti-Stokes fields as quantified by the two-mode principal squeezing variance $
\lambda_{SA} $ behave similarly [see Fig.~\ref{fig3}(d)]. Nevertheless, the
fields entanglement in the exponential regime develops and it becomes stronger
with the increasing pump amplitude $ |\alpha_L^{\rm n}| $. The increase of the
entanglement with the increasing pump amplitude $ |\alpha_L^{\rm n}| $ is
quantified by the logarithmic negativity $ E_{{\cal N},SA} $ [see
Fig.~\ref{fig3}(e)] and the non-classicality depth $ \tau_{SA} $ [see
Fig.~\ref{fig3}(f)]. Whereas the logarithmic negativity $ E_{{\cal N},SA} $
directly quantifies the entanglement, the asymptotic value of the
non-classicality $ \tau_{SA} = 1/2 $ reached for $ |\alpha_L^{\rm n}|
\rightarrow \infty $ identifies the asymptotic state as the most nonclassical
among all two-mode Gaussian states. The values of the logarithmic negativity $
E_{{\cal N},SA} $ and the non-classicality depth $ \tau_{SA} $ attained in the
oscillatory regime, that are maximal for the pump amplitude $
|\alpha_{L,1}^{\rm n}| $, are smaller than the asymptotic values reached in the
exponential regime. This occurs despite the fact that the Stokes and
anti-Stokes photons are perfectly paired for $ |\alpha_{L,1}^{\rm n}| $ and it
is caused by smaller fields intensities in the oscillatory regime.

Steering shows asymmetry in both regimes [see Fig.~\ref{fig3}(g,h)]. Owing to
the microscopic mechanism of the Raman process, steering of the anti-Stokes
mode by the Stokes mode is more efficient than the opposed case. The only
exception occurs in the oscillatory regime for $ |\alpha_{L,1}^{\rm n}| $ where
the Stokes and anti-Stokes field properties are identical and the steering is
symmetric. Similarly as in case of the entanglement measures, the steering
parameters $ \mathcal{S}_{S\rightarrow A} $ and $ \mathcal{S}_{A\rightarrow S}
$ attain larger values in the exponential regime as a consequence of greater
fields intensities for sufficiently high pump amplitudes not shown in
Fig.~\ref{fig3}(g,h). Whereas the states exhibiting the nonlocal Bell
correlations ($ {\cal B}_{SA} > 2 $) are observed in the periodic regime for
the pump amplitudes around $ |\alpha_{L,m}^{\rm n}| $, $ m=1,2,\ldots $ [see
Fig.~\ref{fig3}(i)], the Bell correlations cannot be obtained in the
exponential regime. To get the curves in Fig.~\ref{fig3}(i) [and also
Fig.~\ref{fig6}(i) below], we have numerically optimized the parameters of the
Bell measurement [$ \beta_{S1}= -\beta_{A1} = i \sqrt{\cal J}$, $ {\cal J} =
3.5\times 10^{-3} $, $ \beta_{S2}= -\beta_{A2} = -q\beta_{S1}$, $ q = 3.09 $,
for the symbols, see Eq.~(\ref{eq:non-max}) below] to reach the Bell parameter
$ \mathcal{B}_{\max}=2.313 $. We note that the phases of the optimized complex
parameters differ from those appropriate for parametric down-conversion to
compensate for the negative sign of the coefficient $ D_{SA} $.

According to the above discussion, optimal conditions for pairing of the Stokes
and anti-Stokes photons are found in the oscillatory regime for the pump
amplitudes $ |\alpha_{L,m}^{\rm n}| = (2m-1)\pi /\sqrt{\epsilon-1} $, $
m=1,2,\ldots $. At these $ |\alpha_{L,m}^{\rm n}| $, the mean numbers of Stokes
and anti-Stokes photons coincide and they solely depend on the ratio $ \epsilon
$:
\begin{equation} 
 \langle\hat{n}_S \rangle^{{\rm id},m} = \langle\hat{n}_A \rangle^{{\rm id},m}
   = 4\epsilon/ (\epsilon-1)^{2}.
\label{eq:int-max}
\end{equation}
We also have $ D_{SA}^{{\rm id},m} = -2\sqrt{\epsilon}(\epsilon+1)
/(\epsilon-1)^{2} $. Both the mean photon numbers $ \langle\hat{n}_S
\rangle^{{\rm id},m} $ and $ \langle\hat{n}_A\rangle^{{\rm id},m} $ as well as
the modulus of coefficient $ D_{SA}^{{\rm id},m} $ decrease with the increasing
ratio $ \epsilon $. Physically, this means that the increasing strength of the
nonlinear anti-Stokes interaction relative to the Stokes one, that makes the
population oscillations in the anti-Stokes interaction faster, effectively
decouples the anti-Stokes mode and, moreover, disturbs the Stokes interaction.
We note that for the pump amplitudes $ |\tilde{\alpha}_{L,m}^{\rm n}| = 2m\pi
/\sqrt{\epsilon-1} $ and $ m=1,\ldots $ the Stokes and anti-Stokes modes (as
well as the vibrational mode) return to the initial vacuum states.

Returning back to the condition $ \langle\hat{n}_S \rangle^{{\rm id},m} =
\langle\hat{n}_A\rangle^{{\rm id},m} $, the conservation law of photon/phonon
numbers gives the mean phonon number $ \langle\hat{n}_V \rangle^{{\rm id},m} $
equal to zero, i.e., the vibrational mode is in the vacuum state and the common
state of the Stokes and anti-Stokes modes is pure. We also have $ |D_{SA}^{{\rm
id},m}|^2 = B_{S}^{{\rm id},m} [B_{S}^{{\rm id},m}+1] $. This relation for the
coefficients in the normal characteristic function $ C_{\cal N} $ is well known
from parametric down-conversion, in which it describes ideal twin beams with
perfect correlations in the signal and idler photon numbers [for details, see
Appendix \ref{sec:app}]. However, the coefficient $ D_{SA}^{{\rm id},m} $ has
the negative sign contrary to the case of parametric down-conversion. This
means, that, on average, the positive fluctuation $ \Delta\hat{A}_S $ of the
Stokes-field amplitude is accompanied by the negative fluctuation $
\Delta\hat{A}_A $ of the anti-Stokes-field amplitude. This behavior originates
in the form of the anti-Stokes interaction in which the creation of a phonon is
accompanied by the annihilation of an anti-Stokes photon. In analogy to the
state generated in parametric down-conversion \cite{Mandel1995}, the common
state of the Stokes and anti-Stokes fields is written as (for details, see
Appendix~B):
\begin{eqnarray}   
 |\psi\rangle_{SA}^{\rm id} &=& \sum_{n_S,n_A=0}^\infty (-1)^{n_S} \sqrt{p_{SA}^{\rm id}(n_S,n_A)} |n_S\rangle_S
  |n_A\rangle_A, \nonumber \\
 & & \hspace{-2mm} p_{SA}^{\rm id}(n_S,n_A) = \delta_{n_S,n_A} B^{n_S}/ (1+B)^{1+n_S} ,
\label{33}
\end{eqnarray}
where $ |n_S\rangle_S $ ($ |n_A\rangle_A $) means the Fock state with $ n_S $
($ n_A $) Stokes (anti-Stokes) photons, $ p_{SA}(n_S,n_A) $ stands for the
joint Stokes--anti-Stokes photon-number distribution and $ B $ gives the mean number of photon pairs.

More generally, the joint Stokes--anti-Stokes photon-number distribution $
p_{SA}^{\rm id} $ has nonzero probabilities only for $ n_S \ge n_A $ provided
that $ n_V = 0 $ \cite{perina2011joint}:
\begin{eqnarray}  
  p_{SA}^{\rm id}(n_{S},n_{A})&=& \left( \begin{array}{c} n_S \\ n_A \end{array} \right)
  \frac{\left(B_{A}^{\rm id}\right)^{n_{A}}\left(B_{S}^{\rm id}-B_{A}^{\rm id}\right)^{n_{S}-n_{A}}}{\left(1+B_{S}^{\rm id}\right)^{1+n_{S}}}, \nonumber \\
  & & \hspace{5mm} n_{S}\ge n_{A}.\label{eq:pnlv}
\end{eqnarray}
The ideally-paired joint photon-number distribution $ p_{SA}^{\rm id} $ is
compared with the more general one in Eq.~(\ref{eq:pnlv}) in
Figs.~\ref{fig3}(a,b).
\begin{figure}  
  \includegraphics[width=0.99\hsize]{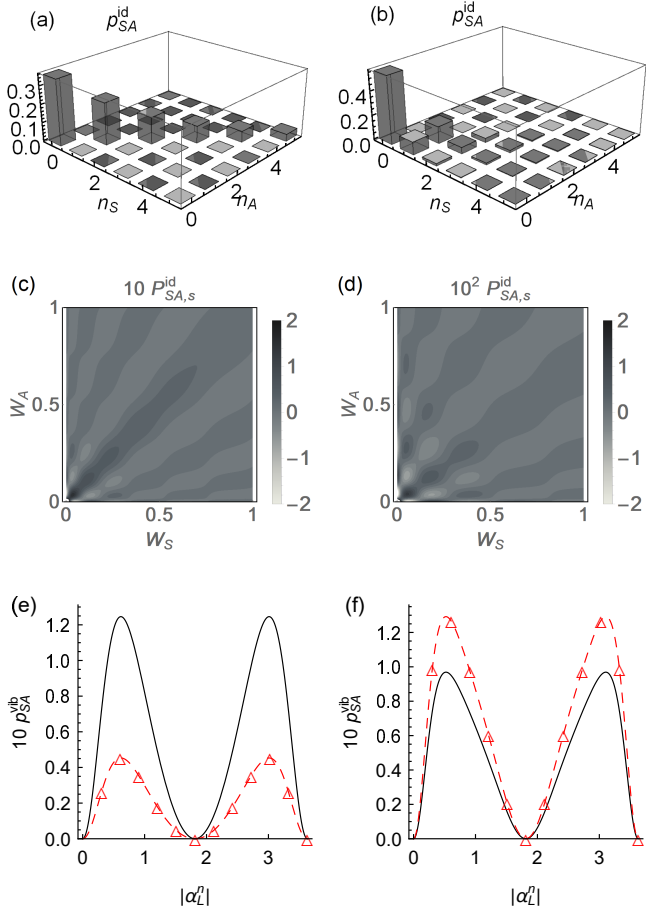}

 \caption{(a,b) Joint Stokes and anti-Stokes photon-number distribution $ p_{SA}^{\rm id}(n_{S},n_{A})$
  and (c,d) the corresponding quasi-distribution $ P_{SA,s}^{\rm id}(W_{S},W_{A}) $ of integrated intensities
  for (a,c) $ s=0.12 $ and pump amplitude $ |\alpha_{L,1}^{\rm n}|$ and (b,d) $ s=0.45 $ and
  $ |\alpha_{L,1}^{\rm n}|/2$; $ \epsilon = 4 $, $ \gamma^{\rm n} = 0 $, $ n_V = n_T = 0 $.
  In (e) and (f), probabilities $p_{SA}^{\rm vib}(1,0)$ [black plain curves] and $p_{SA}^{\rm vib}(0,1)$ [red curves with $\triangle$] are plotted
  assuming $ n_V=0.1 $ (e) and $ n_V=0.5 $ (f).}
\label{fig4}
\end{figure}

The $s$-ordered quasi-distribution $ P_{SA,s}^{\rm id} $ of integrated Stokes
($ W_S $) and anti-Stokes ($ W_A $) intensities belonging to the photon-number
distribution $ p_{SA}^{\rm id} $ and determined along the formula
\cite{perina2011joint}
\begin{eqnarray} 
 P_{SA,s}^{\rm id}(W_{S},W_{A})&=& \frac{4}{(1-s)^2} \exp\left(-\frac{2(W_{S}+W_{A})}{1-s}\right)
 \nonumber \\
 & & \hspace{-12mm} \times \sum_{n_{S},n_{A} =0}^{\infty}  \frac{p^{\rm id}_{SA}(n_{S},n_{A})}{n_{S}!n_{A}!}
  \left(\frac{s+1}{s-1}\right)^{n_{S}+n_{A}}  \nonumber \\
 & & \hspace{-12mm} \times L_{n_{S}}\left(\frac{4W_{S}}{1-s^2}\right)
   L_{n_{A}}\left(\frac{4W_{A}}{1-s^2}\right)
\label{32}
\end{eqnarray}
documents the nonclassical character of the analyzed fields via its negative
values. In Eq.~(\ref{32}), the symbol $ L_k $ stands for the Laguerre
polynomials~\cite{Morse1953}. To demonstrate the properties of the $ s
$-ordered quasi-distributions $ P_{SA,s}^{\rm id} $, we have determined them
for the balanced [$ \langle \hat{n}_S\rangle = \langle \hat{n}_A\rangle $, for
the joint Stokes--anti-Stokes photon-number distribution $ p_{SA}^{\rm id} $,
see Fig.~\ref{fig4}(a)] as well as un-balanced [$ \langle \hat{n}_S\rangle \neq
\langle \hat{n}_A\rangle $, for $ p_{SA}^{\rm id} $, see Fig.~\ref{fig4}(b)]
Stokes and anti-Stokes fields. The areas with negative values of the
corresponding quasi-distributions $ P_{SA,s}^{\rm id} $ are of the triangular
form, as shown in Figs.~\ref{fig4}(c,d). If the mean photon numbers of the
Stokes and anti-Stokes modes are unbalanced, the negative triangular areas get
a characteristic tilt [compare the graphs in Figs.~\ref{fig4}(c) and
\ref{fig4}(d)].

\begin{figure}  
  \includegraphics[width=0.99\hsize]{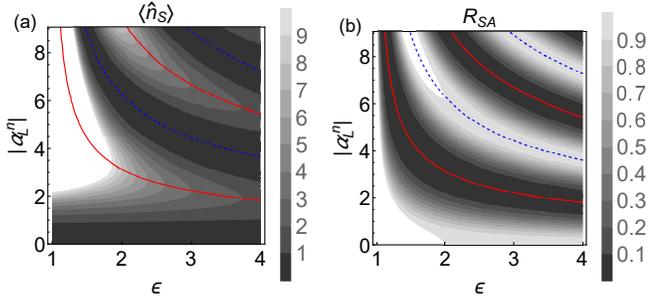}

 \caption{(a) Stokes mean photon number $ \langle \hat{n}_S \rangle $
  and (b) noise-reduction-factor $ R_{SA} $ as they depend on ratio $ \epsilon $ and normalized
  pump amplitude $ |\alpha_{L}^{\rm n}| $. The red solid [blue dashed] curves
  are given as $ |\alpha_{L,m}^{\rm n}| = (2m-1)\pi / \sqrt{\epsilon-1} $
  [$ |\tilde{\alpha}_{L,m}^{\rm n}| = 2m\pi /\sqrt{\epsilon-1} $], $ m=1,\ldots $,
  and identify local maxima (minima) in $ \langle \hat{n}_S \rangle $ and
  minima (maxima) in $ R_{SA} $. The used parameters are: $ \epsilon = 4 $,
  $ \gamma^{\rm n} = 0 $, $ n_V =
  n_T = 0 $.}
\label{fig5}
\end{figure}

According to Eq.~(\ref{eq:int-max}), the fields intensities at $
|\alpha_{L,m}^{\rm n}| $, $ m=1, \ldots $, are not bounded for the ratio $
\epsilon = 1 $, i.e., when the Stokes and anti-Stokes coupling constants equal.
The fields intensities monotonically decrease with the increasing ratio $
\epsilon $ [see Fig.~\ref{fig5}(a)]. The ratio $ \epsilon $ also influences the
other fields parameters, as described by the following formulas:
\begin{eqnarray}  
  g_{SA}^{(2) {\rm id},m} &=& (\epsilon^{2}+6\epsilon+1) / (4\epsilon), \nonumber \\
  R_{SA}^{{\rm id},m} &=& 0, \nonumber\\
  \lambda_{SA}^{{\rm id},m} &=& (\sqrt{\epsilon}-1)^2 / (\sqrt{\epsilon}+1)^{2}, \nonumber \\
  E_{\mathcal{N},SA}^{{\rm id},m} &=& {\rm max} \bigl[0, -\ln\left( \lambda_{SA}^{{\rm id},m} \right)\bigr] , \nonumber \\
  \tau_{SA}^{{\rm id},m} &=& {\rm max} \bigl[0, \left(1-\lambda_{SA}^{{\rm id},m} \right) / 2 \bigr], \nonumber \\
  \mathcal{S}_{S\rightarrow A}^{{\rm id},m} &=& \mathcal{S}_{A\rightarrow S}^{{\rm id},m} =
    {\rm max} \bigl\{0, \ln\left[ (\epsilon^{2}+6\epsilon+1) / (\epsilon-1)^{2}\right]
    \bigr\}, \nonumber \\
  \mathcal{B}^{{\rm id},m}_{SA} &= & \max_{q,\mathcal{J}} \Biggl\{
   \exp\left[-\frac{4 \mathcal{J}}{\lambda_{SA}^{{\rm id},m} }\right] -
   \exp\left[-\frac{4 q^2 \mathcal{J}}{\lambda_{SA}^{{\rm id},m}}\right] \nonumber\\
  && \hspace{-14mm}+2 \exp\left[-\frac{2 \mathcal{J} \left[(q^2+1)(\epsilon^2+6 \epsilon +1) -8 q \sqrt{\epsilon}(\epsilon +1)\right]
    }{(\epsilon -1)^2}\right] \Biggr\}. \nonumber \\
  & &
\label{eq:non-max}
\end{eqnarray}
Whereas the Stokes and anti-Stokes photons are perfectly paired ($ R_{SA}^{{\rm
id},m} = 0 $) for any fields intensity [see Fig.~\ref{fig5}(b)], the quantum
phase correlations are ideal for $ \epsilon = 1 $ ($ \lambda_{SA}^{{\rm id},m}
= 0 $) and they monotonically decrease with the increasing ratio $ \epsilon $.
The phase correlations become classical ($ \lambda_{SA}^{{\rm id},m} = 1 $) in
the limit $ \epsilon \rightarrow \infty $. The entanglement between the Stokes
and anti-Stokes fields, quantified either by the logarithmic negativity $
E_{\mathcal{N},SA}^{{\rm id},m} $ or the non-classicality depth $
\tau_{SA}^{{\rm id},m} $, behave similarly and they disappear in this limit.
The Stokes--anti-Stokes fields are maximally entangled for $ \epsilon = 1 $,
they gradually loose their entanglement with the increasing $ \epsilon $ and,
finally, they are classical for $ \epsilon \rightarrow \infty $. The intensity
correlation function $ g_{SA}^{(2) {\rm id},m} $, that quantifies the violation
of the Cauchy-Schwarz inequality, behaves as a function of $ \epsilon $ in the
opposed way. It does not indicate the Cauchy-Schwarz inequality violation for $
\epsilon = 1 $ ($ g_{SA}^{(2) {\rm id},m} = 2 $), but its value monotonically
increases with the increasing $ \epsilon $ which indicates the non-classicality
of the fields. On the other hand, the noise-reduction-factor $ R_{SA}^{{\rm
id},m} $ equals zero for any $ \epsilon $ owing to perfect pairing of the
Stokes and anti-Stokes photons. The steering in the Stokes--anti-Stokes coupled
system is symmetric, as we have $ \mathcal{S}_{S\rightarrow A}^{{\rm id},m} =
\mathcal{S}_{A\rightarrow B}^{{\rm id},m} $. According to
Eq.~(\ref{eq:non-max}), the greater the fields intensities are, the more
pronounced the steering is. The behavior of intensity correlation function $
g_{SA}^{(2) {\rm id},m} $ and noise-reduction-factor $ R_{SA}^{{\rm id},m} $ in
the limit $ \epsilon \rightarrow \infty $ in which they indicate the
non-classicality, originates in the fact that the fields photon numbers $
\langle\hat{n}_S \rangle^{{\rm id},m} = \langle\hat{n}_A \rangle^{{\rm id},m} $
tend to zero in this limit and they occur in the denominators of fractions
giving the discussed quantities. We note that such behavior is also observed in
parametric down-conversion when it develops from the vacuum state. This means
that these parameters are not suitable quantities for the characterization of
the field properties for very low photon numbers.

\section{The Raman process with an ideal initially populated vibrational mode}\label{sec:nzero-nv}

In the Raman process, there exist two mechanisms that influence the ideal
process discussed in the previous section. The vibrational mode is in a thermal
state with a nonzero mean phonon number and its dynamics suffers from the
losses and the related additional noise.

\begin{figure}  
  \includegraphics[width=0.9\hsize]{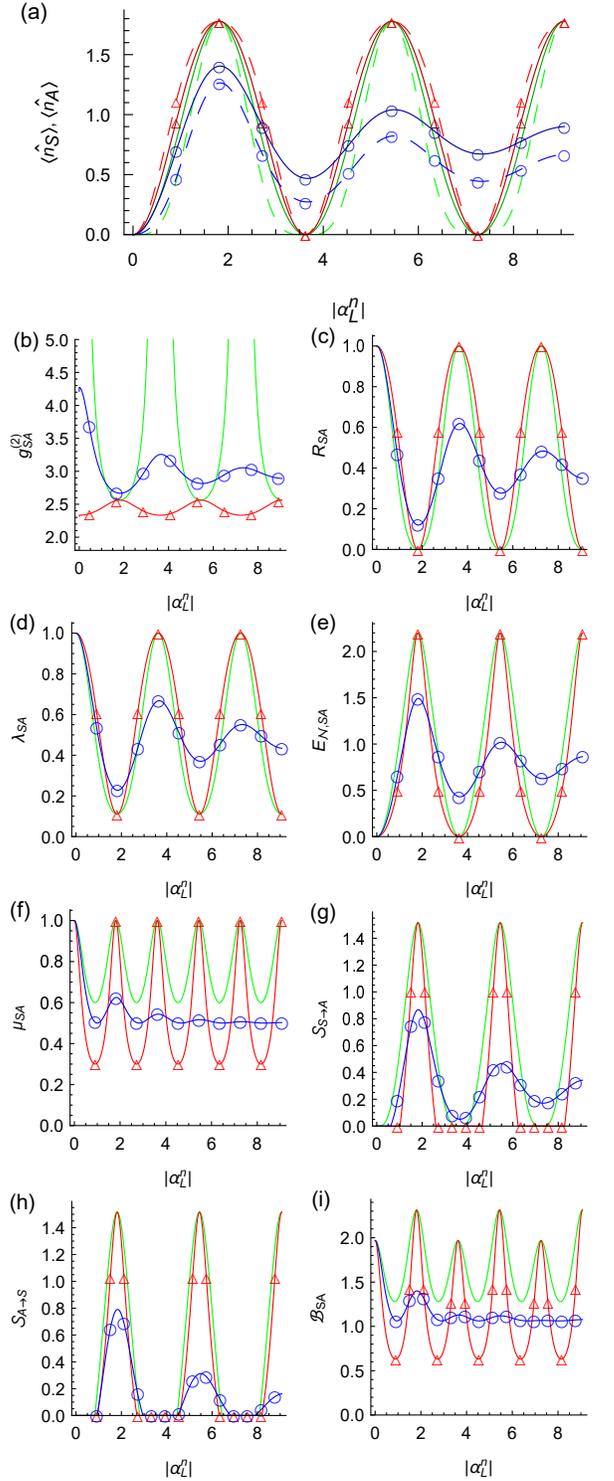}

 \caption{(a) Stokes ($ \langle \hat{n}_S \rangle $, dark solid curves) and anti-Stokes
  ($ \langle \hat{n}_A \rangle $, light dashed curves) mean photon numbers, (b) $ g_{SA}^{(2)} $ intensity function,
  (c) noise-reduction-factor $ R_{SA} $, (d) principal squeezing variance $ \lambda_{SA} $, (e) logarithmic
  negativity $ E_{{\cal N},SA} $, (f) purity
  $ \mu_{SA} $, (g,h) steering parameters
  $ \mathcal{S}_{S\rightarrow A} $ (g) and $ \mathcal{S}_{A\rightarrow S} $
  (h) and (i) the Bell parameter $\mathcal{B}_{SA} $ as they depend on normalized pump amplitude $ |\alpha_{L}^{\rm n}|$.
  The used parameters are: $ \epsilon = 4 $; $ n_V =
  n_T = 0 $, $ \gamma^{\rm n} = 0 $ (green plain curves); $ n_V
  = 0.5 $, $ \gamma^{\rm n} = 0 $, $ n_T= 0 $ (red curves with $ \triangle $);
  $ n_V = n_T = 0.1 $, $ \gamma /(\tilde{g}_S|\alpha_L|) = 1 $
  (blue curves with $ \circ $).}
\label{fig6}
\end{figure}

The initial mean phonon number $ n_V $ increases with the temperature of the
Raman medium and it is given by the Bose-Einstein formula. Such phonons then
allow the emission of anti-Stokes photons without the preceding emission of a
phonon in the Stokes interaction. Such anti-Stokes photons thus have no
correlation to the Stokes photons and they weaken the entanglement between the
Stokes and anti-Stokes fields. Using the solution in Eqs.~(\ref{eq:exact-sol}),
we can derive the following expressions for the coefficients of the normal
characteristic function $ {\cal C}_{\cal N} $ and mean vibrational phonon
number $ \langle \hat{n}_V\rangle (z)= B_{V}(z) $ in the oscillatory regime,
\begin{eqnarray}  
 B_{A}^{\rm vib,p}(z) &=& B_{A}^{\rm id,p}(z) + \epsilon n_V \beta(z),
  \nonumber \\
 B_{S}^{\rm vib,p}(z) &=& B_{S}^{\rm id,p}(z) + n_V \beta(z),
  \nonumber \\
 D_{SA}^{\rm vib,p}(z) &=& D_{SA}^{\rm id,p}(z) - 2\sqrt{\epsilon}  n_V
   \beta(z),  \nonumber \\
 B_{V}^{\rm vib,p}(z) &=& n_V \cos^{2}\left(g_{S}z\sqrt{\epsilon-1} \right)
   +  \beta(z),
   \nonumber \\
 & & \beta(z) = \sin^{2}\left(g_{S}z\sqrt{\epsilon-1} \right)/\left(\epsilon-1\right),
\label{37}
\end{eqnarray}
using the coefficients in Eqs.~(\ref{eq:noise-lossless}). According to
{Eqs.~(\ref{37}), the thermal vibrational phonons increase the mean number $
\langle \hat{n}_S\rangle $ of Stokes photons owing to the stimulated emission
in the Stokes interaction accompanied by the generation of additional phonons
[see Fig.~\ref{fig6}(a)]. The thermal vibrational phonons are also responsible
for the generation of additional anti-Stokes photons, but this is accompanied
by the annihilation of phonons. Whereas the mean number $ \langle
\hat{n}_A\rangle $ of anti-Stokes photons increases owing to the presence of
vibrational thermal phonons, the mean number of vibrational phonons $ \langle
\hat{n}_V\rangle(z) $ may increase or decrease as the Raman process proceeds.
Detailed analysis of the formula in Eq.~(\ref{37}) reveals that for $ n_V
<1/(\epsilon-1) $ [$ n_V>1/(\epsilon-1) $] the number of vibrational phonons $
\langle \hat{n}_V\rangle(z) $ first increases [decreases] and then it decreases
[increases]. This behavior originates in the fact that the effective strength
of the Stokes interaction quantified by the probability $ p_{SA}(1,0) $ is
greater than that of the anti-Stokes interaction quantified by the probability
$ p_{SA}(0,1) $ for $  n_V <1/(\epsilon-1) $. On the other hand, the
anti-Stokes interaction is stronger than the Stokes one for $
n_V>1/(\epsilon-1) $, as documented in Figs.~\ref{fig4}(e,f).

The greatest values of the Stokes mean photon number $ \langle \hat{n}_S\rangle
$ as well as the logarithmic negativity $ E_{{\cal N},SA} $ are found for the
pump amplitudes $ |\alpha_{L,m}^{\rm n}| = (2m-1)\pi /\sqrt{\epsilon-1} $, $
m=1,2,\ldots $, in accordance with the ideal case. Also here, the mean number $
\langle \hat{n}_A\rangle $ of anti-Stokes photons equals that of the Stokes
photons. A bit surprisingly, the greatest Stokes and anti-Stokes mean photon
numbers $ \langle \hat{n}_S\rangle^{m} $ and $ \langle \hat{n}_A\rangle^{m} $,
respectively, are equal and coincide with those of the ideal case. The mean
phonon number $ \langle \hat{n}_V\rangle^{m} $} for this case equals the
initial one, as it follows from the conservation law of photon/phonon numbers
in Eq.~(\ref{5}). This property originates in the general form of the system
dynamics without phonon damping that predicts periodicity in the evolution of
the vibrational mode two times faster than that of the Stokes and anti-Stokes
modes [see Eqs.~(\ref{eq:noise-lossless}) and (\ref{37})]. This means that when
the vibrational mode returns for the first time to its initial state, the Stokes
and anti-Stokes modes are left in certain nontrivial state. The conservation of
the photon/phonon numbers as expressed in Eq.~(\ref{5}) then gives the same
mean photon numbers for both Stokes and anti-Stokes modes.

The thermal phonons drive the Stokes--anti-Stokes dynamics towards greater
mixedness [see the purity $ \mu_{SA} $ in Fig.~\ref{fig6}(f)] first, but the
combined Stokes--anti-Stokes field returns back into a pure state later. This
results in considerably nonclassical values of the $ g_{SA}^{(2),m} $ intensity
function [see Fig.~\ref{fig6}(b)], noise-reduction-factor $ R_{SA}^{m} $
[Fig.~\ref{fig6}(c)], two-mode principal squeezing variance $ \lambda_{SA}^{m}
$ [Fig.~\ref{fig6}(d)] and logarithmic negativity $ E_{{\cal N},SA}^{m} $
[Fig.~\ref{fig6}(e)]. Nonzero mean phonon numbers $ n_V $ also cause the loss
of the ability to steer both the Stokes and anti-Stokes modes for small pump
amplitudes $ |\alpha_{L}^{\rm n}| $ [see Figs.~\ref{fig6}(g,h)]. This ability
is restored for larger values of the pump amplitude $ |\alpha_{L}^{\rm n}| $.
The steering by the Stokes mode appears for smaller values of the pump
amplitude $ |\alpha_{L}^{\rm n}| $ compared to that by the anti-Stokes mode.
However, and the most importantly, for the pump amplitudes $ |\alpha_{L,m}^{\rm
n}| $, for which the Stokes and anti-Stokes modes are equally populated, the
combined Stokes--anti-Stokes state is pure and identical to that in
Eq.~(\ref{33}) found in the ideal case ($ n_V = 0$). Also, for the pump
amplitudes $ |\alpha_{L,m}^{\rm n}| $ the Bell parameter $ {\cal B}_{SA} $
reaches its maximal value independently on the value of $ n_V $ [see
Fig.~\ref{fig6}(i)].

We note that, using the formulas (\ref{eq:noise-lossless}) and {(\ref{37}), the
ratio $ {\cal R} \equiv \langle \hat{n}_A\rangle / \langle \hat{n}_S\rangle $
of the anti-Stokes and Stokes mean photon numbers can be written as a Taylor
expansion in the pump intensity $ |\alpha_L^{\rm n}|^2 $:
\begin{equation}   
  {\cal R} = \frac{B_{A}}{B_{S}} \approx \frac{\epsilon n_V}{n_V+1} +
   \frac{\epsilon}{4\left(n_V+1\right)} \left(1-\frac{\epsilon n_V}{n_V+1}\right) |\alpha_L^{\rm n}|^2.
\label{38}
\end{equation}
According to Eq.~(\ref{38}) the experimental determination of the ratio $ {\cal
R} $ as it depends on the pump intensity $ |\alpha_L^{\rm n}|^2 $ gives us the
ratio $ \epsilon $ of the anti-Stokes and Stokes coupling constants as well as
the mean number $ n_V $ of thermal phonons. If the constants $ r_a $ and $ r_b
$ in the experimental function $ {\cal R} = r_a + r_b|\alpha_L^{\rm n}|^2 $ are
known, we have $ \epsilon = r_a + 4r_b/(1-r_a) $ and $ n_V = r_a(1-r_a)/(4r_b)
$.

The Stokes and anti-Stokes fields also exhibit spatially nonlocal correlations
in their intensities. They are characterized by the correlation function  $
\langle \Delta I_S(z_S) \Delta I_A(z_A)\rangle $ of the Stokes [$ \Delta
I_S(z_S) $] and anti-Stokes [$ \Delta I_A(z_A) $] intensity fluctuations at
positions $ z_S $ and $ z_A $, respectively, that is derived in the form:
\begin{equation} 
 \langle \Delta I_S(z_S) \Delta I_A(z_A)\rangle = |d_{SA}(z_S,z_A)|^2.
\label{39}
\end{equation}
The formula for the coefficient $ d_{SA}(z_S,z_A) $ in Eq.~(\ref{10})
considered for the zero-temperature reservoir ($ n_T = 0 $) expresses this
correlation as an interference of two factorized terms. Whereas only the second
term is nonzero for $ n_V = 0 $, the  first term dominates the second one for
large $ n_V $. In both limiting cases, the correlation function $ \langle
\Delta I_S(z_S) \Delta I_A(z_A)\rangle $ is factorized as a function of the
positions $ z_S $ and $ z_A $. This leads to specific factorized forms of the
correlation function $ \langle \Delta I_S(z_S) \Delta I_A(z_A)\rangle $, as
shown in Figs.~\ref{fig7}(a,c). Factorization of the correlation function means
that, from the point of view of this correlation function, the Stokes and
anti-Stokes fields are independent. For nonzero mean vibrational phonon numbers
$ n_V $, the Stokes and anti-Stokes intensities are correlated and they form a
typical tilted interference pattern, as documented in Fig.~\ref{fig7}(b). Both
types of the interference patterns are 2D periodic. However, if the vibrational
mode is damped, the 2D periodicity is lost. In this case, effective damping in
the anti-Stokes mode is stronger than that in the Stokes mode, as evidenced in
Fig.~\ref{fig7}(d).
\begin{figure}  
  \includegraphics[width=0.99\hsize]{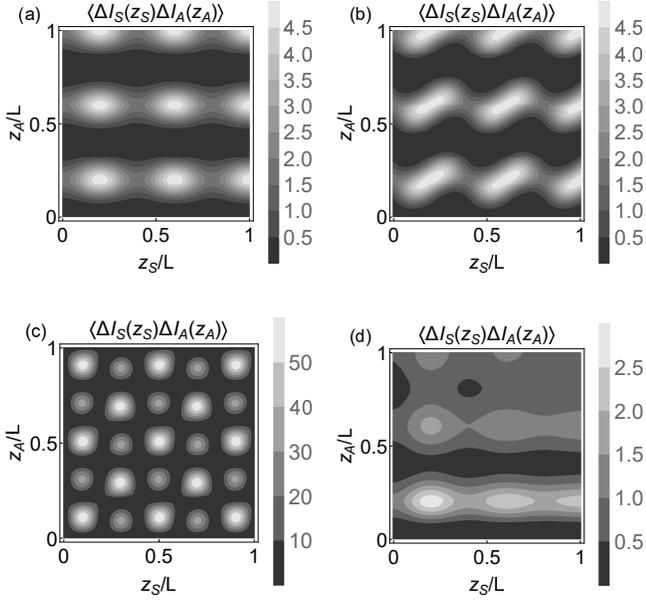}

 \caption{Correlation function $ \langle \Delta I_S(z_S) \Delta I_A(z_A)\rangle $
  as it depends on positions $ z_S $ and $ z_A $ in the Stokes and anti-Stokes
  fields for (a) $ n_V =
   0 $, $ \gamma^{\rm n} = 0 $, (b) $ n_V =1$, $ \gamma^{\rm n} = 0 $, (c) $ n_V =10 $, $ \gamma^{\rm n} = 0 $, (d) $ n_V =
   0 $, $ \gamma /(\tilde{g}_S|\alpha_L|) = 1 $; $ \epsilon = 4 $, $ n_T = 0 $ and
  $ \tilde{g}_S|\alpha_L| = 5\pi /\sqrt{\epsilon-1} $.}
\label{fig7}
\end{figure}

\section{The Raman process with a populated and damped vibrational mode}\label{sec:nzero-nt}

Contrary to the influence of the thermal mean phonon number $ n_V $, damping of
the vibrational mode and the accompanying noise (the fluctuation-dissipation
theorem) cause irreversible changes in the dynamics of the Stokes and
anti-Stokes fields. The above discussed periodic dependence of the analyzed
quantities on the pump amplitude $ |\alpha_{L}^{\rm n}| $ is replaced by that
with fixed asymptotic values in the limiting case $ |\alpha_{L}^{\rm n}|
\rightarrow \infty $ [see Fig.~\ref{fig6}]. Damping in the vibrational mode
decreases the maximal attainable mean photon numbers $ \langle \hat{n}_A\rangle
$ and $ \langle \hat{n}_S\rangle $ [see Fig.~\ref{fig6}(a)] and weakens quantum
correlations between the Stokes and anti-Stokes fields [see the logarithmic
negativity $ E_{{\cal N},SA} $ in Fig.~\ref{fig6}(e)]. Damped oscillations
still occur in the mean photon numbers $ \langle \hat{n}_A\rangle $ and $
\langle \hat{n}_S\rangle $ and logarithmic negativity $ E_{{\cal N},SA} $
considered as a function of the pump amplitude $ |\alpha_{L}^{\rm n}| $. Their
period is larger compared to the ideal case and it increases with the
increasing damping constant $ \gamma $, as it follows from the analytical
formulas in Eqs.~(\ref{eq:exact-sol}) that give the corresponding frequency $
\omega = \sqrt{ |g_A|^2 - |g_S|^2 -\gamma^2} $. With the increasing pump
amplitude $ |\alpha_{L}^{\rm n}| $, the common state of the Stokes and
anti-Stokes modes loses its purity $ \mu_{SA} $ and reaches its asymptotic
value fast [see Fig.~\ref{fig6}(f)]. Damping of the vibrational mode also
reduces the non-classicality of the combined Stokes-anti-Stokes field observed
via the $ g_{SA}^{(2)} $ intensity function [see Fig.~\ref{fig6}(b)],
noise-reduction-factor $ R_{SA} $ [Fig.~\ref{fig6}(c)] and two-mode principal
squeezing variance $ \lambda_{SA} $ [Fig.~\ref{fig6}(d)]. Moreover, it causes
large unbalance in the steering property between the Stokes and anti-Stokes
modes, as documented in Figs.~\ref{fig6}(g,h). Especially, the ability of the
anti-Stokes mode to steer the Stokes mode is considerably reduced. Further, the
Bell violation is only observed for small values of $\gamma^{\rm n}$; it is not
observed for the value of $\gamma^{\rm n}$ used in Fig.~\ref{fig6}(i).

In the asymptotic limit $ |\alpha_{L}^{\rm n}| \rightarrow \infty $, the
coefficients in the normal characteristic function $ C_{\cal N} $ attain the
form:
\begin{eqnarray}  
 B_{A}^{\rm asym} &=& \left(\epsilon n'_T + \epsilon \right)/
  (\epsilon -1)^2, \nonumber \\
 B_{S}^{\rm asym} &=& \left( n'_T +2\epsilon-1 \right) /
  (\epsilon -1)^2,\nonumber \\
 D_{SA}^{\rm asym} &=& -\sqrt{\epsilon} \left( n'_T +\epsilon \right) /
  (\epsilon -1)^2.
\label{40}
\end{eqnarray}
and $ n'_T \equiv (\epsilon -1) n_T $.

Then, the purity $ \mu_{SA}^{\rm asym} $ is derived in the form:
\begin{equation}  
 \mu_{SA}^{\rm asym} = \frac{(\epsilon -1)^2}{(\epsilon +1)(\epsilon-1+2n'_T )}.
\label{41}
\end{equation}
If $ \epsilon = 1 $, i.e., the Stokes and anti-Stokes interactions are
balanced, the asymptotic state is maximally mixed. The increasing ratio $
\epsilon $, that effectively partially decouples the anti-Stokes mode, improves
the purity of the asymptotic state. On the other hand, the greater the mean
reservoir phonon number $ n_T $ is the worse the purity of the asymptotic state
is.

The entanglement of the asymptotic Stokes and anti-Stokes fields quantified
either by the logarithmic negativity $ E_{{\cal N},SA}^{\rm asym} $ or the
non-classicality depth $ \tau_{SA}^{\rm asym} $,
\begin{eqnarray}   
 E_{{\cal N},SA}^{\rm asym} &=& {\rm max} \bigl\{0, \ln \left[ (s_1 - 4s_2 \sqrt{s_3})/2
  \right]/2 \bigr\} , \nonumber \\
 \tau_{SA}^{\rm asym} &=& {\rm max}\left\{ 0, (1-s_2+\sqrt{s_3})/2 \right\},
\label{42} \\
 s_1 &=& 2(\epsilon+1) \bigl[(\epsilon^3+5\epsilon^2-3\epsilon+1) +
  2(\epsilon^2+4\epsilon-1)n'_T \nonumber \\
 & & + 2(\epsilon+1) n_T^{\prime2} \bigr]/ (\epsilon-1)^4 ,\nonumber \\
 s_2 &=& 4(\epsilon+1) (\epsilon+ n'_T ) / (\epsilon-1)^2 ,\nonumber \\
 s_3 &=& \bigl[(4\epsilon^3+\epsilon^2-2\epsilon+1) +
  2(\epsilon+ 1)(3\epsilon-1) n'_T \nonumber \\
 & & + (\epsilon+1)^2
  n_T^{\prime2} \bigr]/ (\epsilon-1)^4 ,\nonumber
\end{eqnarray}
weakens with the increasing ratio $ \epsilon $ as well as the increasing mean
reservoir phonon number $ n_T $. Detailed analysis of the formulas (\ref{42})
reveals that all asymptotic states are entangled for $ \epsilon > 1 $ and $ n_T
\ge 0 $.

The quantum correlations of the asymptotic states can directly be monitored via
the $ g_{SA}^{(2),\rm asym} $ intensity function, noise-reduction-factor $
R_{SA}^{\rm asym} $ and principal squeezing variance $ \lambda_{SA}^{\rm asym}
$:
\begin{eqnarray}  
 g_{SA}^{(2),\rm asym} &=& \frac{\epsilon^2 +2\epsilon -1 + 4\epsilon n'_T
  + 2 n_T^{\prime2} }{ (n'_T+1)(n'_T + 2\epsilon-1) }, \nonumber \\
 R_{SA}^{\rm asym} &=& \frac{ \epsilon + (\epsilon-1)   n'_T +n_T^{\prime2}
   }{ 3\epsilon -1 + (\epsilon+1) n'_T },
  \nonumber \\
 \lambda_{SA}^{\rm asym} &=& \bigl[ \epsilon + n'_T
  \bigr] / \bigl( \sqrt{\epsilon}+1\bigr)^2 .
\label{43}
\end{eqnarray}
Whereas the $ g_{SA}^{(2),\rm asym} $ function indicates the non-classicality
for $ \epsilon > 1 $ ($ g_{SA}^{(2),\rm asym}>2 $), nonclassical values of the
noise-reduction-factor $ R_{SA}^{\rm asym} $ occur only for $ n_T\in \langle 0,
(\sqrt{2\epsilon}+1)/(\epsilon-1) ) $ ($ R_{SA}^{\rm asym} < 1 $) and,
similarly, nonclassical values of the principal squeezing variance $
\lambda_{SA}^{\rm asym} $ are observed only when  $ n_T \in \langle 0,
(2\sqrt{\epsilon}+1)/(\epsilon-1) ) $ ($ \lambda_{SA}^{\rm asym} < 1 $). It
holds for all three quantities, that the greater the mean reservoir phonon
number $ n_T $ is, the less nonclassical their values are. Also, the greater the
ratio $ \epsilon $ is the less nonclassical the values of  $ \lambda_{SA}^{\rm
asym} $ are.

Provided that $ n_T \in \langle 0, 1/(\epsilon-1) ) $ steering occurs in the
asymptotic state as the analysis of the underlying formulas reveals:
\begin{eqnarray}  
 {\cal S}_{S\rightarrow A}^{\rm asym} &=& {\rm max} \left\{0,
  \ln\left[ \frac{\epsilon^2 + 2\epsilon -1 +2n'_T }{
  (\epsilon+1)(\epsilon -1 +2n'_T) }\right]\right\} ,
  \nonumber \\
 {\cal S}_{A\rightarrow S}^{\rm asym} &=& {\rm max} \left\{0,
  \ln\left[ \frac{\epsilon^2 + 1 +2\epsilon n'_T }{
  (\epsilon+1)(\epsilon -1 +2n'_T) }\right]\right\}.
   \nonumber \\
  & &
\label{44}
\end{eqnarray}
The Stokes mode steers the anti-Stokes mode and vice versa in this case though
the steering of the anti-Stokes mode is stronger than the steering of the
Stokes mode. The increasing mean reservoir phonon number $ n_T $ weakens the
ability to steer. The Bell nonlocality was not observed in the discussed steady
state.

The ratio $ {\cal R}^{\rm asym} \equiv \langle \hat{n}_A\rangle^{\rm asym} /
\langle \hat{n}_S\rangle^{\rm asym} $ of the anti-Stokes and Stokes mean photon
numbers derived for the asymptotic state as
\begin{equation}  
 {\cal R}^{\rm asym} = \frac{ B_{A}^{\rm asym} }{ B_{S}^{\rm asym} } = \frac{ \epsilon +
 \epsilon (\epsilon -1) n_T }{ 2\epsilon -1 + (\epsilon -1) n_T }
\label{45}
\end{equation}
can conveniently be used to experimentally determine the ratio $ \epsilon $. If
the mean number $ n_T $ of reservoir phonons can be neglected, we have
\begin{equation}   
 \epsilon =  {\cal R}^{\rm asym} / (2 {\cal R}^{\rm asym} -1 ) .
\label{46}
\end{equation}
Provided that $ n_T> 0 $ we obtain a quadratic equation for $ \epsilon $.

We note that, in the real Raman process, the Stokes and anti-Stokes fields may
suffer from additional scattering, e.g., on the Raman crystal edges and
imperfections. This scattering which means an additional dissipation in the
system can be taken into account similarly as we have done for the vibrational
mode \cite{perina1991quantum}. This leads to weakening of the entanglement and
other non-classical properties of the Stokes and anti-Stokes fields.

{ At the end, we mention generalization of the developed model towards real
experimental conditions. The developed model is stationary. It has been
formulated for monochromatic plane waves whose harmonic dependence along the $
z $ axis is modified by the nonlinear Stokes and anti-Stokes interactions. A
more realistic model should involve non-stationary fields, both in time and the
transverse plane of the interacting fields. This can be done by considering
spectrally and spatially multi-mode fields. In the simplest approximation, we
may model such fields by independent quadruples of spatio-spectral modes in the
laser, Stokes, anti-Stokes, and vibrational modes, similarly as it was done in
\cite{PerinaJr2019a} for parametric down-conversion. The results obtained for
the developed model can then be applied for each quadruple of modes that differ
by their material parameters ($ \epsilon $, $ |\alpha_{L,m}^{\rm n}| $). As the
optimal normalized pump amplitudes $ |\alpha_{L,m}^{\rm n}| $ of quadruples
differ, not all the modes constituting the Stokes and anti-Stokes fields can
simultaneously be under the conditions of ideal photon pairing. This results in
weakening the nonclassical properties of the Stokes and anti-Stokes fields. The
loss of non-classicality depends on detailed conditions of the Raman process.
Here, to illustrate the loss of non-classicality we consider the behavior of
the noise-reduction-factor $ R_{SA}^{\rm m} $ defined in Eq.~(\ref{16}) and
written for multi-mode fields as
\begin{equation} 
 R_{SA}^{\rm m} = \frac{ \langle (\Delta [ \hat{N}_S-\hat{N}_A ] )^2
  \rangle }{ \langle\hat{N}_S\rangle + \langle\hat{N}_A\rangle }
\label{47}
\end{equation}
where $ \hat{N}_b = \sum_{j=1}^{M_b} \hat{n}_{b,j} $ gives the overall
photon-number operator in field $ b $ composed of $ M_b $ modes, $ b= S,A $. In
Fig.~\ref{fig8}(a), we quantify the loss of ideal pairing of the Stokes and
anti-Stokes photons with the increasing number of modes described by parameter
$ \delta $ that defines constant nonzero density of modes in the neighborhood
of the optimal normalized pump amplitude $ |\alpha_{L,1}^{\rm n}| $ defined as
$ |\alpha_{L}^{\rm n}| \in |\alpha_{L,1}^{\rm n}| \langle 1-\delta/2,1+\delta/2
\rangle $. The curve in Fig.~\ref{fig8}(a) for the Raman process with the ideal
vibrational mode in the initial vacuum state was drawn along the following
formula derived from Eqs.~(\ref{eq:noise-lossless}):
\begin{equation} 
 R_{SA}^{\rm m}=\frac{2 \pi  \delta  (4 \epsilon -1)-8 \epsilon
  \sin (\pi  \delta )+\sin (2 \pi  \delta )}{8 \left[\pi  \delta
  (7 \epsilon -1)+16 \epsilon  \sin \left(\frac{\pi  \delta }{2}\right)+(\epsilon +1) \sin (\pi  \delta )\right]}.
\end{equation}
According to the curves in Fig.~\ref{fig8}(a), the greater the number of modes
linearly proportional to the parameter $ \delta $ is, the greater the loss of
non-classicality is. Also, the presence of thermal vibrational phonons results
in the faster loss of non-classicality with the increasing number of modes:
Assuming the value of parameter $ \epsilon $ fixed, the greater the number $
n_V $ of initial thermal phonons is, the greater the value of the
noise-reduction-factor $ R_{SA}^{\rm m} $ is [see Fig.~\ref{fig8}(b)].
\begin{figure}  
   \includegraphics[width=1.0\hsize]{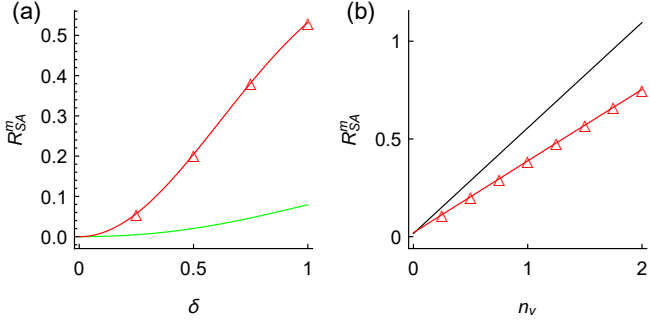}

 \caption{Noise-reduction-factor $ R_{SA}^{\rm m} $ as it depends on (a) the number
  of modes described by parameter $ \delta $ and and (b) initial mean phonon number $ n_V $. Constant density of modes is assumed for
  $ |\alpha_{L}^{\rm n}| \in |\alpha_{L,1}^{\rm n}| \langle 1-\delta/2,1+\delta/2
   \rangle $; (a) $ \epsilon = 4 $, $ \gamma^{\rm n} = 0 $, $ n_V =
   n_T = 0 $ (green plain curve), $ n_V  = 0.5 $ (red curve with $ \triangle $) and
   (b) $ \gamma^{\rm n} = 0 $, $ n_T = 0 $, $ \delta=1/2$, $ \epsilon = 2 $  (black plain curve),
   $ \epsilon = 4 $ (red curve with $ \triangle $).}
\label{fig8}
\end{figure}   }

\section{Conclusions}\label{sec:conc}

The ability of the Raman process to form nonclassical correlations between the
Stokes and anti-Stokes fields has been analyzed. The quantum model involving
independent Stokes and anti-Stokes interactions has been developed using the
operator solution of the Heisenberg equations. According to the model, if
damping of the vibrational mode in the thermal state is negligible and the
anti-Stokes nonlinear interaction is stronger than the Stokes one, the Stokes
and anti-Stokes fields are composed of only photon pairs for specific laser
pump amplitudes. In the intensity properties, the generated state is equivalent
to the state of an ideal twin beam with the same intensity. The state is thus
highly entangled, exhibits strong sub-shot-noise correlations in the difference
of the Stokes and anti-Stokes photon numbers, two-mode phase squeezing,
steering and the Bell nonlocal correlations. The needed pump amplitudes depend
on the length of the medium and nonlinear coupling constants. The mean Stokes
and anti-Stokes photon numbers in this state are independent of the mean number
of thermal vibrational phonons, but they vary with the ratio of the anti-Stokes
and Stokes coupling constants.

When damping of the vibrational mode is considered as well as the pump
amplitude differs from the above ideal ones, the states of the combined Stokes
and anti-Stokes fields become mixed, but they still exhibit nonclassical
features though these are weaker compared to those of the ideal twin beams.
Asymmetry between the Stokes and anti-Stokes fields belongs to the most
important features of these states: The Stokes field has stronger influence to
the anti-Stokes field than the anti-Stokes field influences the Stokes field.

The obtained results elucidate the important role of the anti-Stokes field and
its correlations with the Stokes field in the Raman process. This may find
application in more detailed characterization of the vibrational mode, which is
the essence of the Raman spectroscopy. Moreover, under suitable conditions the
Raman process generates ideally paired optical fields that are commonly used in
quantum metrology and various quantum-information protocols. Compared to
parametric down-conversion usually used in photon-pair generation, the Raman
process is more complex, but also more versatile. This may be useful, e.g., in
quantum metrology.

\section*{Acknowledgements} The authors thank J. Pe\v{r}ina for fruitful
discussions and reading the manuscript. They acknowledge GA \v{C}R (project No.
18-22102S) and support from ERDF/ESF project `Nanotechnologies for Future'
(CZ.02.1.01/0.0/0.0/16\_019/0000754).

\appendix

\section{Spontaneous parametric down-conversion}\label{sec:app}

The process of spontaneous parametric down-conversion pumped by a strong
classical laser beam with amplitude $ |\alpha_L| \exp(i\varphi_L)
\exp(ik_Lz-i\omega_Lt) $ is described by the following momentum operator $
\hat{G}_{\rm spdc}(z) $ \cite{PerinaJr2015a}:
\begin{eqnarray}  
 \hat{G}_{\rm spdc}(z) &=& \hbar k_{s}\hat{a}_{s}^{\dagger}(z)\hat{a}_{s}(z)
  + \hbar k_{i}\hat{a}_{i}^{\dagger}(z)\hat{a}_{i}(z)  \nonumber \\
 & & \hspace{-5mm} + \Bigl[(\hbar \tilde{g} \hat{a}_{s}^{\dagger}(z)\hat{a}_{i}^{\dagger}(z)
 |\alpha_L|\exp\left(ik_L z+i\phi_L\right) +\textrm{H.c.}\Bigr].\nonumber \\
 & &
\label{A1}
\end{eqnarray}
The annihilation (creation) operators $ \hat{a}_{s} $ and $ \hat{a}_{i} $ ($
\hat{a}_{s}^\dagger $ and $ \hat{a}_{i}^\dagger $) are defined in the signal
and idler modes, respectively. Symbol {$ \tilde{g} $} denotes the nonlinear coupling
constant. The frequencies $ \omega_s $ and $ \omega_i $ of the signal and idler
modes, respectively, obey the relation $ \omega_s + \omega_i = \omega_L $. We
also assume the phase-matching conditions for the signal and idler wave vectors
$ k_s $ and $ k_i $, i.e. $ k_s+k_i= k_L $.

The Heisenberg equations derived from the momentum operator $ \hat{G}_{\rm
spdc}(z) $ take the form:
\begin{eqnarray} 
 \frac{d\hat{a}_{s}(z)}{dz} & = & ik_{s}\hat{a}_{s}(z) + g\hat{a}_{i}^{\dagger}(z)\exp\left(ik_{L}z\right),\nonumber \\
 \frac{d\hat{a}_{i}(z)}{dz} & = & ik_{i}\hat{a}_{i}(z) + g\hat{a}_{s}^\dagger(z)\exp\left(ik_{L}z\right)
\label{A2}
\end{eqnarray}
and $ g = i\tilde{g}|\alpha_L|\exp(i\phi_{L}) $. Introducing the operators $
\hat{A}_b(z) \equiv \hat{a}_b(z)\exp(-ik_bz) $, $ b=s,i $, and assuming $
\phi_L = -\pi/2 $ to make the nonlinear coupling constant $ g $ real and
positive, the solution to Eqs.~(\ref{A2}) is expressed as \cite{PerinaJr2015a}:
\begin{eqnarray} 
 \hat{A}_{s}(z) &=& \tilde{f}_1(z) \hat{A}_{s}(0) + \tilde{f}_2(z)
  \hat{A}_{i}^\dagger(0), \nonumber \\
 \hat{A}_{i}(z) &=& \tilde{f}_1(z) \hat{A}_{i}(0) + \tilde{f}_2(z)
  \hat{A}_{s}^\dagger(0),
\label{A3}
\end{eqnarray}
where
\begin{equation} 
 \tilde{f}_1(z) = \cosh(gz),\hspace{2mm} \tilde{f}_2(z) = \sinh(gz).
\label{A4}
\end{equation}

The coefficients in the normal characteristic function $ C_{\cal N} $ in
Eq.~(\ref{8}) attain for this solution and the initial vacuum states the form:
\begin{eqnarray} 
 B_s(z) &=& B_i(z) = \tilde{f}_2^2(z) , \nonumber \\
 D_{si}(z) &=& \tilde{f}_1(z) \tilde{f}_2(z).
\label{A5}
\end{eqnarray}

\section{Statistical operator of the Stokes--anti-Stokes field in the
`balanced' condition}

Here we consider the vibrational mode without damping and being in the initial
vacuum state. Under these conditions, in the oscillatory regime with the pump
amplitudes $ |\alpha_{L,m}^{\rm n}| = (2m-1)\pi /\sqrt{\epsilon-1} $, $
m=1,2,\ldots $, the mean numbers of Stokes and anti-Stokes photons coincide and
we have:
\begin{eqnarray} 
  B^{{\rm id},m} &=& \langle\hat{n}_S \rangle^{{\rm id},m} = \langle\hat{n}_A \rangle^{{\rm id},m}
   = 4\epsilon/ (\epsilon-1)^{2}, \nonumber \\
  D_{SA}^{{\rm id},m} &=& -2\sqrt{\epsilon}(\epsilon+1) /(\epsilon-1)^{2};
\label{B1}
\end{eqnarray}
$ |D_{SA}^{{\rm id},m}|^2 =  B^{{\rm id},m} ( B^{{\rm id},m}+1 ) $. The
anti-normal characteristic function $ C_{\cal A} $ \cite{perina1992quantum} is
then derived in the form:
\begin{eqnarray}  
 C_{\cal A}(\beta_S,\beta_A) &\equiv& \exp[-|\beta_S|^2 -
  |\beta_A|^2] C_{\cal N}(\beta_S,\beta_A) \nonumber \\
 &=& \exp[ -(B^{{\rm id},m}+1) (|\beta_S|^2 + |\beta_A|^2 ) ] \nonumber \\
 & & \times \exp[ D_{SA}^{{\rm id},m} \beta_S^*\beta_A^* + {\rm c.c.} ].
\label{B2}
\end{eqnarray}
Its Fourier transform, the anti-normal distribution function $ P_{\cal A} $
defined as
\begin{eqnarray}  
 P_{\cal A}(\alpha_S,\alpha_A) &=& \frac{1}{\pi^2} \int d^2\beta_S
  d^2\beta_A \, C_{\cal A}(\beta_S,\beta_A) \nonumber \\
 & & \times \exp[ \alpha_S\beta_S^* + \alpha_A\beta_A^* - {\rm c.c.} ]
\label{B3}
\end{eqnarray}
takes the form \cite{perina1992quantum}:
\begin{eqnarray}  
 P_{\cal A}(\alpha_S,\alpha_A) &=& \frac{1}{\pi^2 (B^{{\rm id},m}+1)}
  \exp[ -|\alpha_S|^2 - |\alpha_A|^2 ] \nonumber \\
 & & \hspace{-10mm} \times \exp[ \tilde{D}_{SA}^{{\rm id},m *} \alpha_S\alpha_A]
  \exp[ \tilde{D}_{SA}^{{\rm id},m} \alpha_S^*\alpha_A^*];
\label{B4}
\end{eqnarray}
$ \tilde{D}_{SA}^{{\rm id},m} = \sqrt{B^{{\rm id},m} / (B^{{\rm id},m}+1) } $.

On the other hand, the relation \cite{perina1992quantum}
\begin{equation} 
 P_{\cal A}(\alpha_S,\alpha_A) = \frac{1}{\pi^2} \langle \alpha_S \alpha_A|
  \hat{\varrho}|\alpha_S \alpha_A\rangle
\label{B5}
\end{equation}
allows us to express the anti-normal distribution function $ P_{\cal A} $ in
terms of the elements $ \langle m_S m_A|\hat{\varrho}|n_S n_A\rangle $ of the
statistical operator $ \hat{\varrho} $ in the Fock basis:
\begin{eqnarray}  
 P_{\cal A}(\alpha_S,\alpha_A) &=& \frac{1}{\pi^2}
 \exp[ -|\alpha_S|^2 - |\alpha_A|^2 ] \sum_{m_S,n_S=0}^{\infty} \sum_{m_A,n_A=0}^{\infty}
   \nonumber \\
 & & \hspace{-10mm} \langle m_S m_A|\hat{\varrho}|n_S n_A\rangle
  \frac{\alpha_S^{*m_S} \alpha_S^{n_S} }{ \sqrt{m_S! n_S!} }
  \frac{\alpha_A^{*m_A} \alpha_A^{n_A} }{ \sqrt{m_A! n_A!} }.
\label{B6}
\end{eqnarray}
Expanding the exponentials on the second line of Eq.~({B4}) into their Taylor
series and comparing the expressions in Eqs.~({B4}) and (\ref{B6}), we arrive
at the formula for the elements of the statistical operator $ \hat{\varrho} $:
\begin{eqnarray}  
 & \langle m_S m_A|\hat{\varrho}|n_S n_A\rangle = \delta_{m_S m_A}
  \delta_{n_S n_A} \frac{ [\tilde{D}_{SA}^{{\rm id},m}]^{* m_S}
  [\tilde{D}_{SA}^{{\rm id},m}]^{n_S} }{ B^{{\rm id},m}+1 } & \nonumber \\
 &= \delta_{m_S m_A} \delta_{n_S n_A} (-1)^{m_S+n_S} \frac{
   [B^{{\rm id},m}]^{(m_S+n_S)/2} }{ [B^{{\rm id},m}+1]^{(m_S+n_S)/2+1} }.&
\label{B7}
\end{eqnarray}
The statistical operator $ \hat{\varrho} $ thus describes a pure state written
in Eq.~(\ref{33}) together with the corresponding photon-number distribution $
p_{SA}^{\rm id}(n_S,n_A) $.


%

\end{document}